\documentclass[prb,twocolumn,superscriptaddress]{revtex4}

\usepackage{amsmath}
\usepackage{amsfonts}

\usepackage{graphicx}
\usepackage{times}

\newcommand{\be}{\begin{equation} }
\newcommand{\ee}{\end{equation} }
\newcommand{\ba}{\begin{eqnarray} }
\newcommand{\ea}{\end{eqnarray} }
\newcommand{\n}{\nonumber \\ }

\newcommand{\mac}{\mathcal}

\newcommand{\bv}{\left( \begin{array} }
\newcommand{\ev}{\end{array} \right ) }

\begin{document}

%%%%%%%%%%%%

\title{SU(2) Slave Fermion Solution of the Kitaev
Honeycomb Lattice Model}

\author{F.~J. Burnell}
\affiliation{Rudolf Peierls Centre for Theoretical Physics, University of Oxford, Oxford OX1 3NP, UK}
\affiliation{All Souls College, Oxford, UK}
\author{Chetan Nayak}
\affiliation{Microsoft Research, Station Q, Elings Hall, University of California, Santa Barbara, CA 
93106, USA}
\affiliation{Department of Physics, University of California, Santa Barbara, CA 93106, USA}

\begin{abstract}
We apply the SU(2) slave fermion formalism to the Kitaev
honeycomb lattice model. We show that both 
the Toric Code phase (the A phase) and the gapless phase of this
model (the B phase) can be identified with
$p$-wave superconducting phases of the slave fermions,
with nodal lines which, respectively, do not or do
intersect the Fermi surface.
The non-Abelian Ising anyon phase is a $p+ip$
superconducting phase which occurs when the B phase
is subjected to a gap-opening magnetic field.
We also discuss the transitions between these phases in this language.
\end{abstract}

\maketitle

%%%%%%%%%%%%

\section{Introduction.}

In Ref. \onlinecite{Kitaev06a}, Kitaev introduced the following
remarkable model of $s=1/2$ spins on a honeycomb lattice
\begin{equation}
\label{eqn:Kitaev-honeycomb}
H = -{J_x}\sum_{x-\text{links}} {S^x_j}{S^x_j}
-{J_y}\sum_{y-\text{links}} {S^y_j}{S^y_j}
-{J_z}\sum_{z-\text{links}} {S^x_j}{S^z_j}\, ,
\end{equation}
where the $z$-links are the vertical links on the honeycomb lattice,
and the $x$ and $y$ links are at angles $\pm\pi/3$ from the vertical.
This model is exactly solvable and has a gapped Abelian
topological phase (the `A phase') which is equivalent to the
Toric Code \cite{Kitaev97}. It also has a gapless phase (the `B phase')
which, when subjected to an appropriate time-reversal symmetry-breaking
perturbation, becomes a gapped non-Abelian topological phase
supporting Ising anyons.

This model is one of the rare instances of an exactly solvable
model of a quantum magnet
which does not order in its ground state and, instead,
condenses into a topological phase. As such, it is a useful
testing ground for theoretical techniques, such as slave
fermion representations, which have been
applied to approximately solve models of frustrated magnets
which are not exactly
solvable. Applying these techniques to Eq. \ref{eqn:Kitaev-honeycomb}
can shed light on the physics of this model and, conversely, on the
applicability of these techniques.

Kitaev solved the Hamiltonian (\ref{eqn:Kitaev-honeycomb})
by introducing a fermionization of the spins
in terms of Majorana fermions. By expressing each spin operator
as a product of two Majorana fermions, the spin model
can be described exactly as a model of Majorana fermions
coupled to a $Z_2$ gauge field. In this description the effect of the gauge field
is particularly transparent: the physical correlators are
captured exactly by the fermionic band structure, and the gauge field
serves only to enforce the fact that only gauge-invariant
observables (e.g. products of spins) are physical.

In this paper, we apply a different fermionization procedure,
the SU(2) slave fermion formalism. This
representation requires a different projection to eliminate redundancies in the Hilbert space
compared to Kitaev's representation
in terms of Majorana fermions; therefore, it is interesting
to see how the same low-energy degrees of freedom emerge.
In the SU(2) slave fermion formalism,
the spins are written in terms of standard,
rather than Majorana, fermionic spinons.
The Hamiltonian of Eq. \ref{eqn:Kitaev-honeycomb} is then
expanded about an RVB mean field state.
We show that this is a stable mean-field theory which captures
the the physical correlation functions of the exact ground state
of Eq. \ref{eqn:Kitaev-honeycomb}. We find that the A phase is a $p$-wave
superconducting state of the slave fermions.
The state is fully gapped because the nodes in the order parameter
do not intersect the Fermi surface. The
Majorana fermions of Kitaev's solution appear
as Bogoliubov-de Gennes quasiparticles of the superconducting
state. The B phase is a $p$-wave superconducting
state with gapless excitations at the nodal points. These
excitations form a single Dirac fermion. When the order
parameter develops an $ip$ component,
the Dirac fermion acquires a mass, and the system
goes into an Ising anyon phase.
The transition point between the A phase and the gapless B phase
is an interesting quantum critical point, which we describe in
terms of superconducting order parameters.

By studying the theory of fluctuations
about the mean-field saddle point, we recover the $Z_2$ gauge field as the unbroken gauge symmetry remaining in the superconducting state.
This situates the ground state of the finely-tuned
Hamiltonian (\ref{eqn:Kitaev-honeycomb}) in the broader
context of spin liquid\cite{AffleckMarston,WenPSG,pwa-rvb,pwa-hightc,kalmeyerlaughlin,LaughlinZou,LaughlinZou2,WWZ} and superconducting
phases, and allows us to understand its phase diagram in terms of
these more familiar phases of matter.

\section{SU(2) Slave Fermion Formulation}
\label{sec:fermion-formulation}

\subsection{Slave Fermion Mean-Field Hamiltonian}

Our starting point is the representation of the spin operators in terms of spinful Dirac fermions, first 
discussed in Ref. \onlinecite
{AffleckMarston}.  We thus write the spin operators $\hat{S}_i$, $i=x,y,z$, as:
\be \label{Eq_Spins}
\hat{S}^i = \frac{1}{2} f^\dag_{i \alpha} \sigma^i_{\alpha \beta} f_{i \beta}
\ee
Here, we have introduced the fermion operators
$f_{i \alpha}$, usually called spinons.
For two-spin interactions of the form $\hat{S}_i^\alpha \hat{S}_j ^\beta$, one way to treat the resulting 
Hamiltonian is to use a Hubbard-Stratonovich transformation
to decouple the $4$-fermion interactions, re-expressing them as 
interactions between a bosonic field $\Phi
$ (which lives on a link in the lattice) and a pair of fermion operators on the sites 
$i$ and $j$ bordering this link.  One may then study the mean-field solutions which can be obtained by 
condensing the bosons.  This is 
often a fruitful way to investigate candidate `spin liquid' ground states, in which the spins are strongly 
correlated but have no spatial 
order.

One important caveat in this formulation is that Eq. \ref{Eq_Spins}
gives a faithful representation of 
the Hilbert space only in the 
subspace of fermionic states for which each site is singly occupied.  
%One solution is to Gutzwiller project the fermionic mean-field wave-functions onto the physical 
%Hilbert space, for example using 
%the Monte Carlo methods of Ref. \onlinecite{Gros}.  In the case at hand this projection 
%does not alter the fermionic 
%spectrum, and therefore all physical correlation functions can be obtained exactly from their mean-field values.
%
Thus at each site  ($i$), we must impose the  $3$ (redundant) constraints
\ba \label{Eq_Constrs}
n_{i \uparrow} + n_{i \downarrow} =1 \n
f^{\dag}_{i \uparrow} f^\dag_{i \downarrow} = 0 
\, , \,\,
 \ \ \ f_{i \uparrow} f_{i \downarrow} = 0 \ \ \ .
\ea
   As explained in Ref. \onlinecite{WenPSG},
 when the Hamiltonian preserves SU($2$) spin rotation symmetry, the 
Lagrange multipliers of these constraints
 can be viewed as the temporal component of an $SU(2)$ 
gauge field,  leading to a theory of fermions coupled to a fluctuating gauge field.  (The spatial components of this gauge field are given by the phases of the fermion kinetic terms, which here arise due to condensation of a  bosonic field -- see Appendix \ref{FluctApp}).  Projection would be enforced by
integrating out the gauge fields. In practice, this is typically
done approximately using perturbation theory in the
fermion-gauge field coupling\footnote
{Usually this is done by introducing a 
fermion flavor index $N$, and treating $\frac{1}{N}$ as a
small parameter so that the perturbative 
expansion is controlled, as described in
Ref. \onlinecite{AffleckMarston}}.

Thus the decoupling (\ref{Eq_Spins}) leads to a description
of the spin model as a theory of fermions (spinons) coupled to 
an $SU(2)$ gauge field.  For the Hamiltonian (\ref{eqn:Kitaev-honeycomb}), we will find that the spinons are in a superconducting phase, such that this gauge symmetry is broken down to $Z_2$, and in particular is fully gapped, such that the effect of dynamical gauge-field fluctuations on the fermion band structure is minimal.  We will nonetheless find that this gauge theory is a useful tool to understand the origin of the various topologically ordered phases described in Ref. \onlinecite{Kitaev06a}.

We begin our analysis with the mean-field description of the exact spin-liquid ground state of the Hamiltonian (\ref{eqn:Kitaev-honeycomb}).
In the case of spin-rotationally-invariant Hamiltonians, such as
the Heisenberg model, the Hamiltonian simplifies considerably when
written in terms of the fermions (\ref{Eq_Spins}). In the absence of
spin-rotational symmetry, as in Eq. \ref{eqn:Kitaev-honeycomb},
the Hamiltonian is more complicated. For instance, the Hamiltonian
on $x$-links takes the form.
\begin{multline}
\hat{S}^x_i \hat{S}^x_j =-\frac{1}{4} \left[ f^\dag_{i \uparrow} f^\dag_{j \uparrow} f_{i \downarrow} f_{j 
\downarrow} +  f^\dag_{i \downarrow}  f^\dag_{j \downarrow} f_{i \uparrow}f_{j \uparrow} \right.\\
\left. +  f^\dag_{i \uparrow} f_{j \uparrow} f^\dag_{j \downarrow} f_{i \downarrow}
+  f^\dag_{i \downarrow} f_{j \downarrow} f^\dag_{j \uparrow} f_{i \uparrow} \right ] 
\end{multline}
with similar terms on the $y$-links, as detailed in Appendix \ref{AMFSect}.
(This form is not unique; using the constraints, it can be rewritten
in different forms which are equivalent in the constraint subspace.)
In the Heisenberg model, by contrast, the Hamiltonian on each link
can be written in the form:
\begin{equation*}
\hat{S}^x_i \hat{S}^x_j + \hat{S}^y_i \hat{S}^y_j + \hat{S}^z_i \hat{S}^z_j
= - \frac{1}{2} f^\dag_{i \alpha} f_{j \alpha} f^\dag_{j \beta} f_{i \beta}
+ \frac{1}{4} f^\dag_{i \alpha} f_{i \alpha} f^\dag_{j \beta} f_{j \beta}
\end{equation*}

As a result of the more complex form of the Hamiltonian,
it is necessary to introduce four Hubbard-Stratonovich
fields to decouple the four-fermi interactions.
For example, the Lagrangian on the $x$-links can be written
in the form:
\begin{multline*}
\mac{L}_x = -\frac{8 (|\Phi_1|^2 + | \Phi_2|^2)}{J_x}
 -\frac{8 (|\Theta_1|^2 + | \Theta_2|^2)}{J_x} \\
+ \Phi_1\left ( f^\dag_{i \uparrow} f_{j \uparrow} + 
f^\dag_{i \downarrow} f_{j \downarrow} \right) +
i \Phi_2 \left ( f^\dag_{i \uparrow} f_{j \uparrow} - f^\dag_{i 
\downarrow }f_{j \downarrow} \right)   + \tilde{ h.c.} \\
+  \Theta_1\left ( f^\dag_{i \uparrow} f^\dag_{j \uparrow} 
+ f^\dag_{i \downarrow} f^\dag_{j \downarrow} \right) +
i \Theta_2 \left ( f^\dag_{i \uparrow} f^\dag_{j 
\uparrow} - f^\dag_{i \downarrow }f^\dag_{j \downarrow} \right)
+ \tilde{ h.c.}
\end{multline*}
where $\tilde{ h.c.}$ is the hermitian conjugate with all spin 
directions reversed.
The Lagrangian can be decoupled in a similar manner on
the $y$- and $x$-links as well, as detailed in Appendix \ref{AMFSect}.

Before proceeding, it will be helpful to pick a unit cell
for the honeycomb lattice. We will label the two different
sites with a unit cell by the index $i=1,2$ and different
unit cells by ${\bf R}={n_1}\hat{\bf x}+
{n_2}(\frac{1}{2}\hat{\bf x} + \frac{\sqrt{3}}{2}\hat{\bf y})$.
Then, we denote the fermion fields by $f_{{\bf R} i\sigma}$.
Their Fourier transforms are defined by:
\begin{equation}
f_{{\bf q},i,\sigma} =\frac{1}{\sqrt{N}} \sum_{\vec{R}} e^{i {\bf R}\cdot {\bf q}}
\,f_{{\bf R} i\sigma}
\end{equation}
where $N$ is the total number of lattice sites.

To proceed, we assume that $\Phi_i$, $\Theta_i$
acquire non-zero expectation values.
We parametrize these expectation values by
$t_{ij,\alpha}$, $\Delta_{ij,\alpha}$, $\alpha=\uparrow,\downarrow$,
as explained in Appendix \ref{MFSect}.  Unlike in the case of Heisenberg interactions, 
to describe the Kitaev model we must condense both hopping and superconducting order parameters
or else the mean-field equations will not be satisfied (except
in the special case $J_x = J_y =0$, $J_z\neq 0$),
as shown below. (In the Heisenberg case, hopping and $d$-wave
superconducting terms can be rotated into each other by a
gauge transformation. This is not true for the $p$-wave superconducting
case considered here.)
Because $SU(2)$ spin rotation invariance
is explicitly broken on each link, the latter involve the spin-polarized 
superconducting terms $\Delta_{\uparrow}, \Delta_{\downarrow}$.  Thus,
replacing the fields $\Phi_i$, $\Theta_i$ by their expectation values,
we obtain the mean-field Hamiltonian:
\ba \label{Eq_MarH}
H =& \frac{1}{2}& \!\!\sum_{{\bf q},\sigma} \psi^\dag_{{\bf q} \sigma} \begin{bmatrix} 
0 & t_{ \sigma }({\bf q}) & 0 & \Delta_{ \sigma }({\bf q}) \\
t^*_{ \sigma }({\bf q}) & 0 & -\Delta_{ \sigma }(-{\bf q}) & 0 \\
0 & -\Delta^*_{ \sigma }(-{\bf q}) & 0 & -t^*_{ \sigma }(-{\bf q}) \\
\Delta^*_{ \sigma }({\bf q}) & 0 & -t_{ \sigma }(-{\bf q}) \\
\end{bmatrix}
\psi_{{\bf q} \sigma}
\n
\psi^\dag_{{\bf q}} &= &  \left ( \begin{array}{cccc}
f^\dag_{{\bf q},1, \sigma} & f^\dag_{{\bf q},2, 
\sigma} & f_{-{\bf q},1, \sigma} & f_{-{\bf q},2, \sigma}
\end{array}
\right)  
\ea
(Here the factor of $\frac{1}{2}$ in the first line compensates
for the fact that the expression (\ref{Eq_MarH}) counts each term in the 
Hamiltonian twice. Alternatively, we could sum over half the
Brillouin zone.) If we write $\psi_q$ in components, it has
three indices (in addition to momentum), $\psi_{{\bf q}i\sigma a}$,
where $i=1,2$ is a sublattice index, $\sigma=\uparrow,\downarrow$
is a spin index, and $a=\pm$ is a particle-hole index.

Since we will often be using Pauli matrices to act on these indices,
we will, to avoid confusion, introduce three different notations
for Pauli matrices. We will use $\sigma^{x,y,z}_{\alpha\beta}$
for Pauli matrices acting on spin indices; $\mu^{x,y,z}_{ij}$
for Pauli matrices acting on sublattice indices; and $\tau^{x,y,z}_{ab}$
for Pauli matrices acting on particle-hole indices. (Of course,
it is precisely the same three matrices in all three cases.)

By requiring self-consistency
of the expectation values, we can express $t_{ij,\alpha}$, $\Delta_{ij,\alpha}$
in terms of $J_{x,y,z}$, as shown in Eq. \ref{Eq_MFUp}. 
At the saddle point of interest, the relevant parameters are: 
\ba \label{MFHam}
t_{\uparrow}(q) & = &
 -  \frac{i}{16} \left( e^{i \vec{q}\cdot \hat{l}_1 } J_x+e^{i \vec{q}\cdot \hat{l}_2 } J_y  \right ) \n
 \Delta_{\uparrow}(q) & = &  - \frac{i}{16} \left( e^{i \vec{q}\cdot \hat{l}_2 } 
J_y-e^{i \vec{q}\cdot \hat{l}_1 } J_x   \right ) \n
 t_{\downarrow}(q) &=& - \frac{i}{16} \left( e^{i \vec{q}\cdot \hat{l}_1 } J_x+e^{i \vec{q}\cdot \hat{l}_2 } J_y+2 J_z \right ) \n
\Delta_{\downarrow}(q) &=& \frac{i}{16} \left( e^{i \vec{q}\cdot \hat{l}
_1 } J_x+e^{i \vec{q}\cdot \hat{l}_2 }   J_y \right )
\ea
where $\hat{l}_{1,2} = \frac{\sqrt{3}}{2} \hat{y} \pm \frac{1}{2} \hat{x}$ are the lattice vectors.  

The band energies and eigenfunctions of $H_{MF}$ reveal the correspondence between this picture 
and the Majorana fermion 
decoupling of Ref. \onlinecite{Kitaev06a}.  The mean-field spectrum consists of $3$ flat bands,  with energies:
\ba
\label{eqn:flat-spectra}
\epsilon_{\uparrow x}& =& \pm \frac{ J_x}{8} \ \ \ \ \  \epsilon_{\uparrow y} = \pm\frac
{ J_y}{8} \ \ \ \ \ \epsilon_
{\downarrow z} = \pm \frac{ J_z}{8} 
\ea
and one dispersing band, of energy
\be
\label{eqn:dispersing-spectrum}
\epsilon_{\downarrow}(q) =  \pm\frac{ 1}{8}| J_x e^{ i \vec{q} \cdot \hat{l}_1 } +J_y e^{i 
\vec{q} \cdot \hat{l}_2 } + J_z | \ \ \ .
\ee
(Since we have included an explicit factor of $1/2$ in our definition
of the spin operators $\vec{S}_i$, our $J_{x,y,z}$ are $4$ times
larger than Kitaev's. There is an additional explicit factor of
$4$ in his definition of the spectrum in
Eqs. 31 and 32 in Ref. \onlinecite{Kitaev06a}. This accounts for the
factor $16$ between our spectra.)
The corresponding eigenvectors are naturally
expressed in terms of the Majorana fermions 
\ba \label{Eq_Mar2Dir}
b^y_{{\bf q}i} =  f^\dag_{{\bf q} i \uparrow } + f_{-{\bf q} i \uparrow} &\ \ \ \ \
b^x_{{\bf q} i} = i \left( f^\dag_{{\bf q} i \uparrow } - f_{-{\bf q} i 
\uparrow } \right ) \n
 b^z_{{\bf q} i} =  f^\dag_{{\bf q} i \downarrow} + f_{-{\bf q} i \downarrow i} & \ \ \ \ \ c_{{\bf q} i} = i \left( f^\dag_{{\bf q} i \downarrow } - f_{-{\bf q} i \downarrow } \right ) \ \ \ . 
 \ea
 We have used the same labels as Ref. \onlinecite{Kitaev06a} for these operators.
 
 However, this is not a unique mapping. For instance,
 we could, instead, take $c = -(f^\dagger_\uparrow + f_\uparrow)$,
 $b^x =  i(f^\dagger_\downarrow - f_\downarrow)$,
 $b^y =  f^\dagger_\downarrow + f_\downarrow$,
 $b^z =  i(f^\dagger_\uparrow - f_\uparrow)$.
 Furthermore, the mean-field Hamiltonian has a different expression
 in terms of these operators than in the mean-field theory of
 Ref. \onlinecite{Kitaev06a}. For example, the bilinears $b^z_{R,1} b^z_{R,2}$ 
 do not commute with the mean-field Hamiltonian. The reason
 for this is that if the spin operators are expressed in terms of the
 $f,f^\dagger$s according to Eq. \ref{Eq_Spins}, and then
 the $f,f^\dagger$s are written in terms of $c, {b^x}, {b^y}, {b^z}$,
 according to Eq. \ref{Eq_Mar2Dir}, then we will not obtain the
 same representation as in Ref. \onlinecite{Kitaev06a}.
 Only after the constraints are imposed do the
 operators in Eq. \ref{Eq_Mar2Dir}
 become equivalent to Kitaev's. This is explained in more detail
 in Appendix \ref{sec:Relation-to-Majorana}.

% In this basis, we have: 
%\ba  \label{Eq_MFEvex}
%\psi_{{\bf q}i\uparrow\pm}& =&\frac{1}{2} \left (  b^x_{{\bf q}i}
%\pm i\, b^y_{{\bf q}i} \right ) \cr
%\psi_{{\bf q}i\downarrow\pm}& =&\frac{1}{2} \left (  b^z_{{\bf q}i}
%\pm i\, c_{{\bf q}i} \right )
%\ea
The eigenvectors corresponding to the eigenvalues
(\ref{eqn:flat-spectra}) and (\ref{eqn:dispersing-spectrum}) are
given by:
\ba  \label{Eq_MFEvex}
\alpha_{x\pm}(q)
& =&\frac{1}{2} \left (  i e^{i \vec{q} \cdot \hat{l}_1 } b^x_{q, 1}
\pm b^x_{q,2} \right ) \cr
\alpha_{y\pm}(q) &=& \frac{1}{2} \left (  i e^{i \vec{q} \cdot \hat{l}_2 } b^y_{q, 1} \pm b^y_{q, 2 } \right )\cr
\alpha_{z\pm}(q)&=& \frac{1}{2} \left (   i b^z_{q, 1} \pm b^z_{q, 2}\right )  \cr
\alpha_{0\pm}(q) &=& \frac{1}{2} \left (  i e^{i \theta_q} c_{q,1} \pm c_{q,2}  \right ) 
\ea
where $\theta_q = Arg \left(J_x e^{ i \vec{q} \cdot \hat{l}_1 } +J_y e^{i \vec{q} \cdot \hat{l}_2 } + J_z \right)
$, and in all cases $+$ 
corresponds to the negative-energy solution.
The $b^\alpha_{q,i}$ therefore lie in the $3$ flat bands, and are localized on $x, y$, and $z$ links respectively, and $c$ is the dispersing Majorana mode identified by Ref. \onlinecite{Kitaev06a}.  

Hence the saddle point (\ref{MFHam}) reproduces exactly the description of Ref. \onlinecite{Kitaev06a}, with the precise mapping between the fermions $f_{q, \sigma, i} $ and Kitaev's  Majorana fermions given by Eq. (\ref{Eq_Mar2Dir}).
The only difference is that Ref. \onlinecite{Kitaev06a}
does not include the energy of the flat bands, so that $b^{x,y,z}$  enter only in determining the band structure of the remaining Majorana mode $c$.  The fermionic mean-field energy we obtain per unit cell at half-filling
\be
- \frac{1}{8} \left(J_x +J_y + J_z \right ) - \frac{2}{n_{sites}} \sum_q \epsilon_q
\ee 
However, the first term is cancelled by the zero-point energy arising from terms of the form $\frac{|\Phi_i|^2}{J_{x,y,z} }$,  $\frac{|\Theta_i|^2}{J_{x,y,z} }$ in the Hubbard-Stratonovich Hamiltonian,
%of the bosonic order parameters $\Phi_i$, $\Theta_i$, 
so we
are left with precisely the same energy as in Kitaev's solution.

Superficially, we have obtained an 8-band mean-field theory from a model of spinful fermions on a lattice with a $2$-site unit cell.  Readers might thus justifiably be concerned that we have in fact obtained double the degrees of freedom that we would have expected. However, we have combined $f_{{\bf q}i\sigma}$
and $f^\dagger_{-{\bf q}i\sigma}$ into the same spinor; consequently,
we should restrict ${\bf q}$ to half the Brillouin zone to avoid double-counting.

\subsection{Slave Fermion Band Structure}

To understand the physics of this model, it is useful
to focus on the band structure of the down-spin fermions.
It suffices to consider the case $J_x = J_y =J$:
\ba \label{Eq_MFBand}
{\epsilon_\downarrow}(q) &=& \pm \frac{J}{8} \left \{ \left( \frac{J_z}{J} + 2 \cos \frac{q_x}{2} \cos \frac{\sqrt{3}q_y }{2}   \right )^2 \right . \n
&& \left. + 4 \left ( \cos \frac{q_x}{2} \sin \frac{\sqrt{3}q_y  }{2} \right )^2 \right \}^{1/2} \ \ \ .
\ea
This describes a pair of bands which cross at either
$0$ or $2$ distinct points in the Brillouin zone.
Following Ref. \onlinecite{Kitaev06a}, we will call the former case,
which occurs for $|J_z| > 2 | J|$, the A phase.
In the A phase, the spectrum is fully gapped.
When $|J_z| < 2 | J|$, there are two Majorana cones
in the spectrum or, equivalently, a single Dirac cone.
This is the B phase. Our objective here is to understand
how this band structure arises in the slave fermion
superconductor, and use this analogy to understand the transitions
between these phases.  

We begin with a more scrupulous analysis of the nature of the superconducting state.  Since the character of the phase is determined by the dispersing fermion band, we will focus on the mean-field Hamiltonian for the down spins.
If we combine the down-spin fermions on the two sublattices
into the following spinor,
\begin{equation} \label{Eq_PsiSpinor}
\Psi_q =
\begin{pmatrix}
 f_{q 1 \downarrow} \cr f_{q2\downarrow}
 \end{pmatrix}
 ,
\end{equation}
then the Hamiltonian has the general form:
\ba \label{Eq_Hdd}
H_{\rm down}  &=& 
 \Psi^\dag_{q} \left[ \epsilon_q^{(x)} \mu_x+ \epsilon_q^{(y)} \mu_y \right ]   \Psi_{q} \n
 && +\Psi^\dag_{q}  \left(\Delta^{(s)}_q \mu_y + \Delta^{(t)}_q \mu_x \right ) (\Psi^\dag_{-q})^T + h.c. \n
 && +  \frac{J_z}{8} \left (2 - \frac{J}{J_z}  \right )  \Psi^\dag_{q} \mu_y \Psi_q
\ea
where we have taken $J_x = J_y =J$,
and
\begin{eqnarray}  \label{DnDis}
\epsilon_q^{(x)} &=&  \frac{J}{8} \cos \frac{q_x}{2} \sin \frac{\sqrt{3} q_y}{2}\cr
\epsilon_q^{(y)} &=& \frac{J}{16}(1+ 2 \cos \frac{q_x}{2} \cos \frac{\sqrt{3} q_y}{2})
\end{eqnarray}
represent the kinetic energy for fermions hopping on the honeycomb lattice. The third line corresponds to an in-plane `magnetic field' in pseudospin space due to the enhanced hopping along the $z$ links. This term shifts the positions of the Majorana cones, but is otherwise unremarkable. 

The second line is a superconducting pairing term along the $x$- and $y$-links.
Both 
\ba \label{DnSC}
\Delta^{(s)}_q &=&  \frac{J}{8} \cos \frac{ \sqrt{3} q_y}{2}  \cos \frac{q_x}{2}  \n
\Delta^{(t)}_q &=&- \frac{J}{8}\sin \frac{ \sqrt{3} q_y}{2}  \cos \frac{q_x}{2}
\ea
 are non-vanishing in the mean-field state. The superscipts $(s)$ and $(t)$ refer to the fact
that these are pseudospin-singlet and pseudospin-triplet superconducting
order parameters.

If we linearize about the nodes (we work at the isotropic
point, $J=J_z$, for simplicity), then the Hamiltonian for down-spins
takes the form:
\begin{multline}
\label{Eq_HddEff}
H_{\rm down}  = 
 \Psi^\dag_{p} \left[  -\frac{J\sqrt{3}}{32}  p_y \mu_x+
\frac{J\sqrt{3}}{32}  p_x \mu_y  - \frac{J}{16} \mu_y\right ] \Psi_{p} \\
 - \frac{J}{16}  \Psi^\dag_p  \mu_y (\Psi^\dag_{-p})^T + \text{h.c.}\\
+ \frac{J\sqrt{3}}{32}   \Psi^\dag_p   \left[ p_y 
   \mu_x - p_x \mu_y \right ] (\Psi^\dag_{-p})^T + \text{h.c.}
 \end{multline}
Here, $\vec{p}$ is the distance from the node $(4\pi/3,0)$.
This Hamiltonian has four eigenvalues, the two non-dispersing ones
$\pm J_z/8$, and the two dispersing ones in Eq. \ref{Eq_MFBand}.

It is helpful to isolate the dispersing band (unlike the Hamiltonian
(\ref{Eq_Hdd}), which contains both the dispersing and non-dispersing
down-spin bands). To this end, we form the Dirac fermion
\begin{equation}
\label{eqn:eta-def}
\eta_q = e^{i\pi /4}(c_{q 1} -i c_{q 2})
\end{equation}
The mean-field Hamiltonian for $\eta_q$ is (up to a constant):
\begin{equation}
\label{Eq_H1Band}
\tilde{H} = \frac{1}{2}\sum_{q}\left( \epsilon_q \eta^\dag_q \eta_q 
+ \Delta_q \eta^\dag_{q} \eta^\dag_{-q} + h.c. \right)
\end{equation}
where 
\ba
\epsilon_q &=& \frac{1}{8} \left( J_z+ 2 J \cos \frac{q_x}{2} \cos \frac{\sqrt{3} }{2} q_y \right )\\
\Delta_q &=& \frac{1}{4} J  \cos \frac{q_x}{2} \sin \frac{\sqrt{3}}{2} q_y
\ea

To understand this Hamiltonian better, it is useful to
momentarily imagine that $\Delta_q = 0$ and focus on
$\epsilon_q$. The Hamiltonian now describes spinless fermions
on the honeycomb lattice with dispersion $\epsilon_q$.
First consider $J_z > 2J$. We see that
there is no Fermi surface: $\epsilon_q$ is never equal to zero.
Consider the minimum energy excitation, which occurs
at $\vec{q}=(0,\frac{2\pi}{\sqrt{3}})$ and has energy
$J_z - 2J$. Near the minimum the band is approximately
quadratic. There are no excitations near zero energy
because the effective `Fermi energy' lies below the bottom of the band.
Superconductivity does not change this picture very much,
other than to break U(1) symmetry (which is very important
when we go beyond mean-field).
When superconductivity is turned back on, there are no nodes or
nodal excitations because there is no Fermi surface.

For $J_z< 2J$, there is Fermi surface which surrounds
the point $(0,\frac{2\pi}{\sqrt{3}})$. Strictly speaking,
for the usual Brillouin zone this point sits on its boundary,
so half the Fermi surface encircles
$(0,\frac{2\pi}{\sqrt{3}})$ while the other half encircles the
equivalent point $(0,-\frac{2\pi}{\sqrt{3}})$ which differs by
a reciprocal lattice vector. Of course, we could take a different
unit cell for the reciprocal lattice which only includes
one of these two points; then the Fermi surface will surround this point.
We now restore the superconducting gap $\Delta_q$. This
opens a gap on the Fermi surface, except at the points on
the Fermi surface which intersect the nodal line
$q_y = \frac{2\pi}{\sqrt{3}})$. (The nodal line $q_y = 0$ does not
intersect the Fermi surface, except for the point $(4\pi/3, 0)$,
which is equivalent to $(2\pi/3, 2 \pi/\sqrt(3))$ under translation
by a reciprocal lattice vector.) For $2-{J_z}/J\ll 1$ small, the Fermi
surface is approximately circular. Let us expand momenta 
about $(0,\frac{2\pi}{\sqrt{3}})$, so that
$({q_x},{q_y}) \approx (0,\frac{2\pi}{\sqrt{3}}) + (2{p_x},2{p_y}/\sqrt{3})$.
Then ${\epsilon_p} \approx J({p_x^2}+{p_x^2})-\mu$,
where the `Fermi energy' $\mu$ is given by $\mu=2J-{J_z}$,
and $\Delta_p = J\,p_y$. Thus, the Hamiltonian in the B phase
looks like that of a $p_y$ superconductor, which has nodes at $p_y=0$.
As $J_z$ is decreased and the system moves towards the isotropic
point, the nodes move towards the corners of the Brillouin zone,
eventually reaching the graphene spectrum at the isotropic point.

\section{Mean-Field Phase Diagram in the Absence of
Time-Reversal Symmetry-Breaking Perturbations}

We will now apply the mean-field description outlined in the previous section to understanding the 
phase diagram of (\ref{eqn:Kitaev-honeycomb}) in terms of its fermionic band structure and superconducting gap.
As we shall see, the principle advantage of the spinful mean-field decoupling is that it allows us to better understand the 
system's behavior away from the exactly solvable point -- both in terms of proximate phases, and the fate of physical quantities 
such as the spin-spin correlation functions as we perturb the Hamiltonian (\ref{eqn:Kitaev-honeycomb}).  At the end of this section 
we also describe at mean-field level the nature of the phase transition separating the gapped A phase and gapless B phase.

\subsection{The A Phase}  \label{Sect_APhase}

We begin by studying the A phase, for which $J_z > 2 J $
 and the band structure (\ref{Eq_MFBand}) is fully gapped.
%In the effective 1-band superconducting model (\ref{Eq_H1Band}), the A phase arises when the Fermi surface has shrunk such that it never intersects the nodes $q_y  =0, \pm \frac{ 2 \pi}{\sqrt{3}}$ of the superconducting order parameter.  In fact, this occurs because for $J_z > 2 J$ the  band energies (given by $t_q = J_z + 2 J \cos \frac{q_x}{2} \cos \frac{ \sqrt{3} q_y}{2}$) never change sign.  Hence there is no Fermi surface for the nodes to intersect with. 
In this phase, superconductivity, which couples fermions along the $x$- and $y$- links, competes with dimerization along the $z$-links, as is evident from the $2$-band Hamiltonian (\ref{Eq_Hdd}). In the A phase the dimerization term dominates, leading to a fully gapped band structure.
in the extreme limit $J=0$, $J_z \neq 0$, dimerization leads to a gap, even
in the absence of superconductivity.
(Indeed, many fruitful explorations of the A phase treat it as an effective theory of such interacting dimers\cite{APhaseStuffs,APhase1,APhase2,APhase3}).

As seen at the end of the previous section, we may view the
A phase as a spin-polarized $p$-wave superconductor with chemical potential
which lies below the conduction band. 
One amusing consequence of this is that the topological
order of this phase is, as explained in Ref. \onlinecite{OganesyanSondhi},
that of a $Z_2$ gauge theory.
Its topological nature stems from the fact that, in the condensed phase, the only 
remnant of the interactions between gauge fields and matter is a  `statistical' interaction due to the Berry's phase of $\pi$ accrued by a 
charge if it encircles a vortex of flux $\frac{\hbar}{2e}$\cite{DeWildePropitius}.  
This provides an alternative perspective on the well-documented fact\cite{Kitaev06a, APhaseStuffs} that the 
A phase is smoothly connected 
to the so-called Toric code\cite{Kitaev97} -- a model of Ising spins which realizes 
a topological $Z_2$ gauge theory with matter.   In particular, this
highlights that the topological order
of the A phase is not restricted to the set of exactly solvable Hamiltonians described by (\ref{eqn:Kitaev-honeycomb}),
but is that of a garden-variety $s$-wave superconductor.

If we only cared about the single-particle gap, then we could
close the superconducting gap entirely without closing the
total fermion gap. However, the gauge symmetry of the problem
would not be broken down to $Z_2$ in this case, so there
would be gapless gauge field fluctuations about the mean-field solution.
(In the dimerized limit $J_x = J_y =0$, though the $U(1)$ gauge symmetry is 
unbroken these gapless modes are absent since the gauge field cannot 
propagate).  

%Now let's consider what happens when a weak perturbation
%is added to the Hamiltonian. For instance, we could add
%a magnetic field and/or a Heisenberg term. Since the system
%is fully-gapped, perturbation theory can be used, and the
%effect will be small, so long as the perturbation is weak.

Because the A phase is fully gapped, it is  stable to weak perturbations away from the solvable point discussed here.  
For instance, we could add a weak magnetic field and/ or Heisenberg interaction without changing the qualitative 
features of this phase.  Since the system is fully-gapped, perturbation theory can be used, and the effect will be small, so 
long as the perturbation is weak.  This is in contrast to the B phase which, as we will see, is unstable in the face of 
appropriately chosen perturbations.

\subsection{The Nodal B Phase}   \label{Sect_BPhase}

We now briefly describe the B phase, for which $J_z < 2 J $.
Now (\ref{Eq_H1Band}) is the band structure of a $p$-wave superconductor whose nodes intersect the Fermi surface at two distinct points in
the Brillouin zone.

To simplify the algebra, we will consider 
the symmetric point $J_x =J_y =J_z \equiv J$. 
% This point has an additional, `spin-orbit' type symmetry under simultaneous lattice and 
%spin rotations which simplifies its band structure.
The energies of the dispersing Majorana bands are then exactly those of free fermions in a honeycomb lattice.
%\be
%\epsilon_q = \pm \frac{J}{8} \sqrt{1 + 4 \cos^2 \frac{q_x}{2} + 4 \cos \frac{q_x}{2} \cos \frac{ \sqrt{3} q_y}
%{2} }
%\ee
The spectrum is gapless at the points
$\vec{q} = ( \pm \frac{2 \pi}{3}, \frac{2 \pi} {\sqrt{3}} )$
(and at the equivalent points $( \pm \frac{4 \pi}{3}, 0)$,
$( \frac{2 \pi}{3}, -\frac{2 \pi} {\sqrt{3}} )$, which differ
from the first two by reciprocal lattice vectors).
These nodes account for two distinct cones
in the energy spectrum, as in graphene.
However, unlike in graphene, the band structure (\ref{Eq_MFBand}) is that of a pair of bands of dispersing {\it Majorana} fermions. 
In the vicinity of these nodal points, it is useful to rewrite the Hamiltonian (\ref{Eq_H1Band}) in terms of the spinor
\begin{equation}
\chi_q =
\begin{pmatrix}
\eta_q \cr \eta^\dagger_{-q}
 \end{pmatrix}
 ,
\end{equation}
where $\vec{q}$ is restricted to lie in half of the Brillouin zone
to avoid double-counting, e.g. over ${q_x}>0$.
In terms of this spinor, the Hamiltonian can be written in the form:
\ba \label{Eq_JackReb}
H =  \frac{1}{2} \sum_{q_x >0, q_y}
\chi^\dag_q \left [  \Delta_q \tau_x +   \epsilon_q  \tau_z
\right ]  \chi_q    
\ea
In the vicinity of the nodes (at the isotropic point $J_z = J$),
we can expand $\vec{q} = (\frac{4 \pi}{3}, 0) + ({p_x},{p_y})$
and write
\begin{equation}
\tilde{\chi}_p =
\begin{pmatrix}
\eta^{}_{(\frac{4 \pi}{3}, 0)+\vec{p}} \cr \eta^\dagger_{-(\frac{4 \pi}{3}, 0)-\vec{p}}
 \end{pmatrix}
,
\end{equation}
and $\vec{p}$ now ranges unrestricted over small $\vec{p}$
(e.g. over $|\vec{p}|<\Lambda$, for some cutoff $\Lambda$),
i.e. near the nodes.
Expanding $\epsilon = \frac{ \sqrt{3} J}{16} p_y,
\Delta = \frac{ \sqrt{3} J}{16} p_x$,
we can write:
\ba
H &=&  \sum_{\vec{p}}
\tilde{\chi}^\dag_p \left [  \frac{ \sqrt{3} J}{32} p_y \tau_x +   \frac{ \sqrt{3} J}{32} p_x  \tau_z
\right ] \tilde{\chi}_p    \cr
&=&v\int {d^2}x \, \tilde{\chi}^\dagger \left[ i\partial_y \tau_y + 
 i\partial_x  \tau_z \right] \tilde{\chi}
\ea
with $v=\frac{ \sqrt{3}}{32}J$.
Thus, these two Majorana fermions combine to form a single
Dirac fermion. This Dirac cone is formed by combining the two nodes
of a $p_y$ superconductor.
This single Dirac cone does not violate the usual fermion doubling arguments since the gauge symmetry is broken. We will see presently, 
however, that it is central to the non-Abelian statistics of the gapped B$^*$ phase.

We now consider some of the correlation functions of the
B phase. Since there are gapless excitations, the energy-density
will certainly have power-law correlations. How about the spin-spin
correlation function? At the soluble point, this is short-ranged.
Consider, for instance, the ${S^z}-{S^z}$ correlation.
In terms of the slave fermions
${S_i^z} = ({f_{i\uparrow}^\dagger}{f_{i\uparrow}} - 
{f_{i\downarrow}^\dagger}{f_{i\downarrow}})/2$.
Since up and down-spins decouple,
\begin{equation}
\left\langle {S_i^z} {S_j^z} \right\rangle
= \frac{1}{4}\left\langle {f_{i\uparrow}^\dagger}{f_{i\uparrow}}
{f_{j\uparrow}^\dagger}{f_{j\uparrow}} \right\rangle
+  \frac{1}{4}\left\langle {f_{i\downarrow}^\dagger}{f_{i\downarrow}}
{f_{j\downarrow}^\dagger}{f_{j\downarrow}} \right\rangle
\end{equation}
The first term vanishes since it only involves $b^x$ and $b^y$,
and these create/annihilate fermions in the up-spin flat bands.
Here, $b^x$ and $b^y$ are defined in terms of $f_\downarrow$,
$f^\dagger_\downarrow$ according to Eq. \ref{Eq_Mar2Dir}.
(It is important to remember that, although they play the same role
in our analysis as the operators with the same labels in
Ref. \onlinecite{Kitaev06a}, they are not
identical, in spite of the obvious similarity.)
Thus, we are left with
\begin{eqnarray}
\left\langle {S_i^z} {S_j^z} \right\rangle
&=& \left\langle {f_{i\downarrow}^\dagger}{f_{i\downarrow}}
{f_{j\downarrow}^\dagger}{f_{j\downarrow}} \right\rangle/4\cr
&=& \left(1+ \left\langle i b_i^z c_i  \right\rangle
+ \left\langle i b_j^z c_j  \right\rangle
- \left\langle b_i^z c_i\, b_j^z c_j  \right\rangle
\right)/16\cr
&=& 0
\end{eqnarray}
At the mean-field level, this is a free fermion problem,
so we can evaluate these correlation functions.
The Hamiltonian does not mix $b^z$ with $c$, so
$\left\langle i b_i^z c_i  \right\rangle=0$
and $\left\langle b_i^z c_i\, b_j^z c_j  \right\rangle=
\left\langle b_i^z b_j^z \right\rangle \left\langle c_j c_i \right\rangle$.
Since $b^z$ creates a fermion in a flat,
non-dispersing band, $\left\langle b_i^z b_j^z \right\rangle=0$
unless $i$ and $j$ are the same or neighboring sites.

One of the appealing features of the formalism we use is that correlation 
functions in the presence of small perturbations to   the Hamiltonian 
(\ref{eqn:Kitaev-honeycomb})
can be calculated with relative ease.  
For instance, suppose we consider a weak magnetic field
in the $z$-direction, as in Ref. \onlinecite{Tikhonov10}.
This adds a perturbation to the Hamiltonian:
\be
H_{\rm pert} = \frac{1}{2} h_z {\sum_i}({f_{i\uparrow}^\dagger}{f_{i\uparrow}} - 
{f_{i\downarrow}^\dagger}{f_{i\downarrow}}) 
\ee
For small $h_z$, this perturbation does not spoil the basic
structure of the spectrum: there are still three gapped bands
and one gapless one. The up-spin gapped band will
still be non-dispersing and will be at the same energy,
but the corresponding eigenoperators will mix $b^x$ and $b^y$
(unlike the eigenoperators (\ref{Eq_MFEvex}) in the
unperturbed Hamiltonian). 
The down-spin gapped band will now disperse weakly,
but will remain gapped. However, the eigenoperators
for the down-spin bands will now mix $b^z$ and $c$.
Thus, when we compute the $\left\langle {S_i^z} {S_j^z} \right\rangle$
correlation function, $b^z$ will have a small amplitude,
proportional to $h_z$ for small $h_z$,
to create a dispersing fermion. Thus, this correlation
function will have power-law falloff.

To see this more precisely, we add the magnetic field
term to the down-spin Hamiltonian:
\begin{multline}
\label{eqn:Ham+field}
H_{\rm down}  = 
 \Psi^\dag_{p} \left[  -\frac{J\sqrt{3}}{32}  p_y \mu_x+
\frac{J\sqrt{3}}{32}  p_x \mu_y  - \frac{J}{16} \mu_y\right ] \Psi_{p} \\
+ \frac{J\sqrt{3}}{32}   \Psi^\dag_p   \left[ p_y 
   \mu_x - p_x \mu_y \right ] (\Psi^\dag_{-p})^T + \text{h.c.}
\\
 - \frac{J}{16}  \Psi^\dag_p  \mu_y (\Psi^\dag_{-p})^T + \text{h.c.} 
     -\frac{1}{2}{h_z} \Psi^\dag_{p}\Psi_{p}
 \end{multline}
When we diagonalize this Hamiltonian, we find
a new set of eigenoperators $\tilde{\alpha}_{z\pm}$, $\tilde{\alpha}_{0\pm}$.
The eigenoperator $\tilde{\alpha}_{z+}$ creates a fermion in a weakly-dispersing
gapped band and has short-ranged correlation functions.
The eigenoperator $\tilde{\alpha}_{0+}$ creates a fermion in a gapless
band and has power-law correlation functions.
For small $h_z$ (and, for simplicity, small momentum $k$),
we can express the fermions
$\alpha_{z\pm}= \left (   i b^z_{q, 1} \pm b^z_{q, 2}\right )/2$,
$\alpha_{0\pm} = \left (  i e^{i \theta_q} c_{q,1} \pm c_{q,2}  \right )/2$,
in terms of these new eigenoperators as:
\begin{eqnarray}
\alpha_{z\pm} &=& \tilde{\alpha}_{z\pm} \pm \frac{h_z}{2} \tilde{\alpha}_{0\pm}\cr
\alpha_{0\pm} &=&  \mp \frac{h_z}{2} \tilde{\alpha}_{z\pm} +  \tilde{\alpha}_{0\pm}
\end{eqnarray}
Thus, we now have:
\ba
\left\langle b^z_i b^z_j \right\rangle &=& -\left\langle (\alpha_{z+,i}+\alpha_{z-,i})
(\alpha_{z+,j}+\alpha_{z-,j}) \right\rangle\cr
&=& -\Bigl\langle (\tilde{\alpha}_{z+,i}+\tilde{\alpha}_{z-,i} +
{h_z}(\tilde{\alpha}_{0+,i}-\tilde{\alpha}_{0-,i})/2)\, \times\cr
& & {\hskip 0.5 cm}(\tilde{\alpha}_{z+,j}+\tilde{\alpha}_{z-,j} +
{h_z}(\tilde{\alpha}_{0+,j}-\tilde{\alpha}_{0-,j})/2) \Bigr\rangle\cr
&=& \left\langle \tilde{\alpha}_{z+,i}\tilde{\alpha}_{z+,j}\right\rangle
+  \left\langle \tilde{\alpha}_{z-,i}\tilde{\alpha}_{z-,j}\right\rangle\cr
& & + \mbox{$\frac{h_z^2}{4}$}\left(  \left\langle \tilde{\alpha}_{0+,i}\tilde{\alpha}_{0+,j}\right\rangle
+ \left\langle \tilde{\alpha}_{0-,i}\tilde{\alpha}_{0-,j}\right\rangle\right)
\ea
Here, we have assumed, for the sake of concreteness and simplicity,
that $i$ and $j$ are on the 1 sublattice. From the Hamiltonian
(\ref{eqn:Ham+field}), we have for large separation $|{\bf x}-{\bf y}|$
and to zeroeth order in $h_z$:
\begin{multline}
\left\langle \tilde{\alpha}_{0+,{\bf x}}\tilde{\alpha}_{0+,{\bf y}}\right\rangle
+ \left\langle \tilde{\alpha}_{0-,{\bf x}}\tilde{\alpha}_{0-,{\bf y}}\right\rangle
=\\
\int \frac{d\omega}{2\pi}\frac{{d^2}k}{(2\pi)^2}\,
\frac{\frac{J\sqrt{3}}{16}(k_y + i{k_x}/2)\,\,e^{i {\bf k}\cdot({\bf x}-{\bf y})}}
{\omega^2 - \left(\frac{J\sqrt{3}}{16}\right)^2
({k_x^2} + 4 {k_y^2})}
\end{multline}
Therefore, at long distances,
\be
 \left\langle \tilde{\alpha}_{0\pm,{\bf x}}\tilde{\alpha}_{0\pm,{\bf y}}\right\rangle
 \sim \frac{1}{|{\bf x}-{\bf y}|^2}
\ee
Combined
with the $\left\langle c_i c_j \right\rangle$, which has the same-power-law,
this gives an $\left\langle {S_i^z} {S_j^z} \right\rangle$ correlation
function which falls off as $1/r^4$ in the presence of a small
magnetic field, in agreement with the results of Ref. \onlinecite{Tikhonov10}.
 
In the face of perturbations that are not quadratic in the fermions, such explicit calculations are more difficult in general.  However, as is frequently the case in spin-liquid models\cite{WenPSG}, the structure of the Fermi surface (here a pair of Dirac cones) is protected by symmetries of the mean-field state.  Thus small perturbations which do not break any symmetries of the problem cannot open a gap in the spectrum. 

\subsection{Transition between A and B Phases}
\label{ABTransSect}

As we move within the gapless B phase, from the isotropic
point ${J_x}={J_y}={J_z}$ towards the boundary to the A phase,
the two nodal points move together and, at the phase transition
point, merge. The nodes then annihilate as the phase boundary
is crossed. In this section, we focus on the transition point.

As discussed in Section \ref{sec:fermion-formulation},
the dispersing spin-down band can be rewritten as a model
of spinless fermions with $p_y$ superconducting order,
as in Eq. \ref{Eq_H1Band}. At the boundary between A
and B phases, the Fermi surface has shrunk to a point
because the effective chemical potential is precisely
at the bottom of the band. When the effective chemical
potential is at the bottom of the band,
the spectrum is quadratic in the absence of superconductivity.
Superconductivity with $p_y$ pairing symmetry leaves the
spectrum gapless but makes the spectrum linear in one direction.
We now examine this in more detail.
Expanding about the bottom
of the band $({q_x},{q_y}) \approx (0,\frac{2\pi}{\sqrt{3}})
+ (2{p_x},2{p_y}/\sqrt{3})$, we can write the Hamiltonian
(\ref{Eq_H1Band}) in the form:
\ba
\tilde{H} &=& \frac{1}{2}\sum_{p}\left[ \frac{J}{8} p^2\, \eta^\dag_p \eta_p 
-\frac{J}{4} {p_y}\, (\eta^\dag_{p} \eta^\dag_{-p} - \eta^{}_{p} \eta^{}_{-p})\right]\cr
&=& \frac{1}{2} \sum_{p_x>0,p_y} \chi^T_{-p} \left[ -\frac{J}{4}{p_y}I
-\frac{J}{8} {p^2} i\tau_y\right] \chi_p
\ea
where
\begin{equation}
\chi_p =
\begin{pmatrix}
\eta_p \cr \eta^\dagger_{-p}
 \end{pmatrix}
 ,
\end{equation}
If we go to a Majorana basis,
\begin{equation}
\varphi_p =\frac{1}{\sqrt{2} }
\begin{pmatrix}
\eta_p +  \eta^\dagger_{-p} \cr (\eta_p - \eta^\dagger_{-p})/i
 \end{pmatrix}
 ,
\end{equation}
this can be re-written:
\ba
H &=& \frac{1}{2}\sum_{p_x>0,p_y} \varphi^T_{-p} \left[ -\frac{J}{4}{p_y} \tau_z
-\frac{J}{8} {p^2} \tau_y\right] \varphi_p \cr
&=&\frac{1}{2} \int {d^2}x\, \varphi^T \!
\left[ -\frac{J}{4} i\partial_y \tau_z - \frac{J}{8}  \partial^2 \tau_x \right]\varphi
\ea
Therefore, the low-energy theory can be called a single gapless
Majorana fermion, albeit an anisotropic and non-relativistic one.

\section{Beyond Mean Field Theory}

Thus far, we have found a consistent mean-field solution of (\ref{eqn:Kitaev-honeycomb}) using the fermionization (\ref{Eq_Spins}) which reproduces exactly the Majorana fermion band structure  and  phase diagram
of the exact solution proposed by Ref. \onlinecite{Kitaev06a}.  
We next ask what can be said about its fate upon 
including fluctuations of the various 
bosonic fields.  
The answer is not obvious since, unlike the decoupling used by Ref. \onlinecite{Kitaev06a}, the product 
$b^\alpha_i b^\alpha_{i+1}$ on each link does not commute with the full unprojected fermion Hamiltonian (although it does commute with
the quadratic Hamiltonian $H_{MF}$).  
Here we first establish that these fluctuations do not alter the results of the previous sections.  Second,
we demonstrate that at long wavelengths these bosonic modes lead to precisely the $\mathbb{Z}_2$ 
gauge theory of Ref. \onlinecite{Kitaev06a}.  Together, these facts cement the equivalence 
between the fermionization (\ref{Eq_Spins}) and Kitaev's exact solution.  

The underlying reason for this stability is that the unprojected mean-field wave functions we obtain can be mapped via Eq. (\ref{Eq_Mar2Dir}) onto unprojected wave-functions in the Majorana fermionization of Ref. \onlinecite{Kitaev06a}.    Enforcing the $SU(2)$ gauge constraints to reduce the model back to the physical Hilbert space amounts to two things: first, it eliminates the disticntion between different possible mappings between $f_\sigma, f^\dag_\sigma$ and $b_{x,y,z},c$.  Second, it imposes a condition which is equivalent to the $\mathbb{Z}_2$ constraint required for the fermionization of Ref. \onlinecite{Kitaev06a}.  Thus when expressed in the Majorana basis given by (\ref{Eq_Mar2Dir}), the effect of this projection will be to apply the projector relevant to Kitaev's Majorana fermionization.  In this way, both fermionizations lead to the same wave functions after projection.

\subsection{Symmetries and robustness of the mean-field solution}  \label{SymSect}

First, we will show that for the solvable Hamiltonian (\ref{eqn:Kitaev-honeycomb}), the model's unusually large number of symmetries protect  
the exact fermionic band structure.  The mean-field solution is thus exact, in that it correctly describes all correlators of the physical spin degrees of freedom, in spite of the apparent violence done to the wave-function by Gutzwiller projection.  
 
We begin by listing the symmetries which are relevant to this discussion.
The Hamiltonian (\ref{eqn:Kitaev-honeycomb}) has the following
discrete symmetries 
\be
\mathbf{\hat{C}} \ : \  S_i^{x,y,z} \rightarrow  s_{x,y,z} S_i^{x,y,z}
 \ee
 where the sign $s_{x,y,z} = \pm 1$ can be chosen independently for $x$, $y$, and $z$ spin operators.
In the fermionic description, this leads to two discrete symmetries preserved by the mean-field Hamiltonian:
\ba
\mathbf{\hat{C}} &:& f_{q \sigma i}  \rightarrow
f^\dagger_{-q \sigma i}\cr
\mathbf{\hat{S}}&:& f_{q1\sigma} \rightarrow f_{-q1\sigma},\,
f_{q2\sigma} \rightarrow -f_{-q2\sigma} \ \ \ .
\ea
Here, the `charge conjugation' symmetry $\mathbf{\hat{C}}$ is unitary
(it is is simply $\psi_{qi\sigma a}\rightarrow \left(\tau^x\right)_{ab}\
 \psi_{qi\sigma b}$)
while the `sublattice' symmetry $\mathbf{\hat{S}}$ is an anti-unitary symmetry.
Thus, in the mean-field Hamiltonian, $\mathbf{\hat{C}}$
takes $\Delta_{ij}, t_{ij} \rightarrow \Delta_{ij} , t_{ij}$
while $\mathbf{\hat{S}}$ takes $\Delta_{ij}, t_{ij} \rightarrow \Delta^* _{ij} , t^* _{ij}$.
Quadratic Hamiltonians invariant under $\mathbf{\hat{C}}$ have 
eigenstates which are diagonal in the Majorana basis (\ref{Eq_Mar2Dir}).
These symmetries impose an important restriction on $t_{ij}$ and $\Delta_{ij}$.
$\mathbf{\hat{C}}$ is preserved as 
long as $t_{ij}, \Delta_{ij}$ are purely imaginary.
$\mathbf{\hat{S}}$ is preserved so long as there are
no terms directly coupling fermions on the same sublattice. 

Time-reversal symmetry is also respected by the model
and its mean-field solution:
 \be
 \mathbf{\hat{T}} \ \ : \ \  f_{q, \uparrow } \rightarrow f_{-q, \downarrow}\, ,
 \,\, f_{q, \downarrow } \rightarrow - f_{-q, \uparrow}
 \ee
Single-spin terms (i.e. a magnetic field) and three-spin interactions
break this symmetry.
However, not all $\mathbf{\hat{T}}$-breaking perturbations will open a gap
in the B phase: only those perturbations which break $\mathbf{\hat{S}}$
will open a gap in the spectrum, as we will see below.
For example, the magnetic field discussed in Sect. \ref{Sect_BPhase} breaks $\mathbf{\hat{T}}$ and $\mathbf{\hat{C}}$, but not $\mathbf{\hat{S}}$.  
As shown explicitly above, this does not gap the B phase and indeed results in power-law spin-spin correlations.  

The relation between these symmetries is:
\begin{equation}
\mathbf{\hat{T}} = \mathbf{\hat{S}}  \mathbf{\hat{G}^x}  \mathbf{\hat{C}} 
\end{equation}
where the symmetry $\mathbf{\hat{G}^x}$ is given by: 
 \ba  \label{Eq_Residuals}
\hat{G}^{x,y}: \ \  f_{i \uparrow}  \rightarrow f^\dag_{i \downarrow} \n
 t_{ij, \uparrow}, \Delta_{ij ,\uparrow}  &\rightarrow & t_{ij, \downarrow},  \Delta_{ij, \downarrow} \n
 t_{ij, \downarrow}, \Delta_{ij ,\downarrow}  & \rightarrow&  t_{ij, \uparrow},  \Delta_{ij, \uparrow} \n
 \ea
which are a discrete subset of the off-diagonal $SU(2)$ rotations interchanging up and down spins.  In the mean-field solution these are no longer local symmetries.  However, they remain global symmetries of the theory, whose effect is to rotate between different possible mappings between the four Majorana fermions ($c, b^{x,y,z}$), and the four self-adjoint combinations $f_{i \sigma}^\dag + f_{i \sigma}, i \left( f_{i \sigma}^\dag - f_{i \sigma} \right )$ of the spinful fermions.  Thus $\mathbf{\hat{T}}$ is a {\it projective} symmetry --  a symmetry that maps the system to a different but gauge equivalent saddle point.  Such projective symmetries are important to classifying the phases of spin-liquid systems\cite{WenPSG}.

\begin{center}
\begin{figure}[h!]
\includegraphics[height=2in]{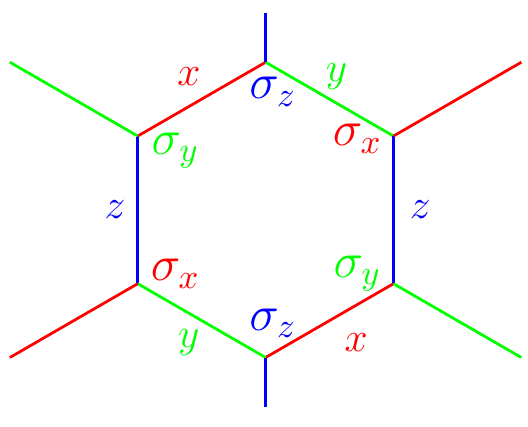}
\caption{ \label{ComFig} [Color Online]  The product of spin operators conserved separately on each plaquette by the 
Kitaev Hamiltonian (\ref
{eqn:Kitaev-honeycomb}).  The Hamiltonian (\ref{eqn:Kitaev-honeycomb}) distinguishes between three types of links on the honeycomb lattice, which we call $x$-, $y$-, and $z$- links ( color-coded red, green, and blue respectively here).  On $x$-links the spin-spin interaction term is $S^{(x)}_i S^{(x)}_j$, and similarly for $y$- and $z$- links.  The product of spin operators shown here -- a product around a plaquette of the spin variable associated with the `external' edge at each vertex -- commutes with the spin Hamiltonian.  }
\end{figure}
\end{center}

Besides these more generic discrete symmetries, 
Eq. (\ref{eqn:Kitaev-honeycomb}) represents a somewhat special point in a more extended space of 
similar spin Hamiltonians:  there is a product of spin operators on each plaquette which commutes with 
$H$.  This is:
\be  \label{Eq_Pcons}
\mac{P} = \prod_{i=1}^6 S^{e(i)}(i)   = \pm \frac{1}{2^6} 
\ee
where $e(i) = z$ for a vertex which sits between $x$ and $y$ links on the plaquette, $y$ for a vertex 
which sits between $x$ and $z$ 
links on a plaquette, and $x$ for a vertex which sits between $y$ and $z$ links on a plaquette (see Fig. 
\ref{ComFig}).   In the ground 
state, the value of this operator is positive on each plaquette\cite{Kitaev06a}.  

In terms of the fermionic operators, $\mac{P}$ can be written as
\be
\mac{P}_{f} \equiv
P_0\left( \prod_{i=1}^6 b^{\alpha}_i b^\alpha_{i+1} \right )
P_0
\ee
where $P_0$ denotes Gutzwiller projection onto singly occupied states, $\alpha =x,y,z$ on $x, y$, and 
$z$-links, respectively, and  $b^
\alpha_i$ are the Majorana fermions defined in Eq. (\ref{Eq_Mar2Dir}).  (Since the quantity in parentheses is not $SU(2)$ gauge invariant, the projection operator is necessary in this case).  
In the mean-field state, each species of Majorana fermion is localized on the appropriate links, with $ 
\langle b^\alpha_i b^\alpha_{i+1} 
\rangle_{MF} = 1/2$.   Terms annihilated by $P_0$ do not contribute, since $ \langle f^\dag_{i \uparrow} 
f^\dag_{i \downarrow} \rangle_
{MF} =\langle f_{i \uparrow} f_{i \downarrow} \rangle_{MF}=0$.  Hence we find that the mean-field value  
\be \label{Eq_PfMF}
\mac{P}_{f} = \langle b^x_1 b^x_2 \rangle  \langle b^y_2 b^y_3 \rangle  \langle b^z_3 b^z_4 \rangle
 \langle b^x_4 b^x_5 \rangle   \langle b^y_5 b^y_6 \rangle  \langle b^z_6 b^z_1 \rangle  = \frac{1}{2^6} 
 \ee
is precisely that of the exact solution.  

We now show that, combined with the discrete symmetries mentioned above, conservation of $\mac{P}_f$ prevents fluctuations about mean-field
from altering the fermionic band structure in any way.
We will first establish that the symmetries forbid any
terms other than those in Eq. (\ref{Eq_PfMF}) from 
contributing to $\mac{P}_f$. If there can be no further
contributions to $\mac{P}_f$ induced by 
fluctuations, however, then also no 
spectral weight can be transferred from the equal-time
correlation functions of the $b^\alpha$, as 
otherwise we would not arrive at the 
correct value for $\mac{P}$. This means that all
further-neighbor correlators must vanish exactly.  

By Wick's theorem, we need only consider the possibility of 
other pairings of the fermionic 
operators which give a non-zero contribution to $\mac{P}_f$.  The only possibility allowed by $\mathbf
{\hat{C}}$ and $\mathbf{\hat{P}}\mathbf{\hat{S}}$ 
is to give a non-vanishing expectation value to terms of the form $\langle b_1^x b_4^x\rangle, \langle 
b_2^y b_5^y \rangle$, etc.  Thus 
we consider:
\be \label{Eq_MFPert}
\langle b^x_1 b^x_4 \rangle  \langle b^y_2 b^y_5 \rangle  \langle b^z_3 b^z_6 \rangle
 \langle b^x_2 b^x_5 \rangle   \langle b^y_3 b^y_6 \rangle  \langle b^z_4 b^z_1 \rangle
\ee
However,  the interacting Hamiltonian for the up spins 
decouples exactly into separate 
Hamiltonians for each chain of $x-y$ links in the lattice.  In particular, the full Hamiltonian contains no 
interaction term coupling $b_1^x
$ and $b_4^x$, as they lie on different chains.  Hence interactions cannot shift $\langle b^x_1 b^x_4 
\rangle$ from its mean-field value 
of $0$.  As (\ref{Eq_MFPert}) is the only extra contribution to $\mac{P}_f$ not explicitly forbidden by 
symmetry, we conclude that Eq. (\ref
{Eq_PfMF}) must remain valid in the full solution, and that consequently no fermion bilinears can be 
shifted from their mean-field 
values.

\subsection{Gauge theory of fluctuations about mean field}
\label{GaugeSect}

Thus far, we have shown how to reproduce Kitaev's mean-field portrait of the exact spin-liquid ground state  using the fermionization (\ref{Eq_Spins}), and argued that including fluctuations about mean-field will not change the fermionic band structure.  Hence we have obtained an alternative mean-field description of the ground state of (\ref{eqn:Kitaev-honeycomb}) which reproduces faithfully the spin correlators of the exact ground state.  

Though the mean-field solutions describe identical physics, however, the fermionization (\ref{Eq_Spins}) differs quite dramatically from that of Ref. \onlinecite{Kitaev06a} in the nature of the bosonic variables, and consequently, the theory of fluctuations about mean-field.  After Hubbard-Stratonovich transforming the $4$-fermion interactions, we obtain bosonic fields which condense to give both the hopping and superconducting order parameters, as well as the $SU(2)$ gauge fields associated with the constraint (\ref{Eq_Constrs}).  One might therefore wonder why these do not lead to significantly different physical theories after fluctuations about mean-field have been accounted for.  Here we address this question, allowing us to posit that (\ref{Eq_MarH}) describes a gapped spin-liquid phase which exists even away from the exactly solvable limit of the Hamiltonian (\ref{eqn:Kitaev-honeycomb}).

The bosonic fluctuations about mean-field can be separated into the following degrees of freedom.  There are three scalar fields describing fluctuations in the amplitudes of the various kinetic and superconducting terms.   All of these are massive, and as we shall see two of them can be interpreted as Higgs fields for the broken $SU(2)$ symmetry.  In addition, there are three independent fields associated with phase fluctuations of the various link variables.   These can be identified as an $SU(2)$ gauge field (describing
phase fluctuations of the spin-symmetric hopping term) and two `Goldstone bosons' associated with the phases of the order parameters breaking the $SU(2)$ symmetry.  
We will briefly discuss each type in turn; a more detailed
analysis is presented in  Appendix \ref{FluctApp}.  

We begin with the scalar fields describing fluctuations in the amplitude of the various bosonic order parameters that fix the mean-field fermionic band structure.
The general form of the Hubbard-Stratonovich action ensures that all of the scalar fields are massive, with energy gaps of order $\frac{1}{J}$ at the isotropic coupling point.  Because of this mass gap, fluctuations in the amplitudes of the mean-field parameters are not generally expected to have an important effect on the fermions.  The notable exception to this\cite{Dimers} is in cases when they destabilize the spin liquid saddle point in favor of a `dimerized' state with spins hopping predominantly along a subset of links in the lattice.  As we discuss in Sect. \ref{Sect_APhase}, an analogue of the dimerized phase does occur for anisotropic $J_{x,y,z}$; in general we may therefore conjecture that away from the solvable point this phase boundary may be shifted, but that fluctuations of the mean-field hopping and superconducting amplitudes will not qualitatively alter the phase diagram.  

Next, we consider the impact of phase fluctuations described
by the $SU(2)$ gauge theory. Naively, the gauge theory is
strongly-fluctuating, since there is no small parameter in the
problem. However, the ground state of (\ref{eqn:Kitaev-honeycomb}) 
is a Higgs phase, so that the gauge field is massive.  (Importantly, this 
explains why the gauge theory is not confined).

To see that the model (\ref{eqn:Kitaev-honeycomb}) is in a Higgs phase, 
we view the mean-field solution (\ref{Eq_MarH}) 
as a condensate of two independent order parameters in the
adjoint representation of $SU(2)$.  As explained in detail in Appendix \ref{FluctApp},  the 
combination of superconducting and spin-antisymmetric hopping terms break the $SU(2)$ 
gauge symmetry.  This leaves only the residual $Z_2$ gauge symmetry group one normally finds in a superconductor:
 \be
 f_{i \sigma},\ \  f^\dag_{i \sigma}  \rightarrow - f_{i \sigma},  -f^\dag_{i \sigma}  \ \   \ \ \  t_{ij, \sigma}, \Delta_{ij ,\sigma}  \rightarrow t_{ij, \sigma},  \Delta_{ij, \sigma} \ \ \ 
 \ee
comprising the residual $Z_2$ symmetry of the $U(1)$ subgroup broken by superconductivity.  
As a result of the Anderson-Higgs phenomenon, the dynamical fluctuations in the gauge field are suppressed at long wavelengths, so that gauge field fluctuations are not expected to substantially alter the fermionic band structure.  (Here the gauge field results from the constraints of the purely $2$ dimensional system, and consequently is fully gapped unlike the electromagnetic gauge field in thin-film superconductors.)
However, the gauge field makes itself felt in the interesting topological
structure of the spin-liquid phase.

An alternative route for a gauge field to acquire a mass is through
the generation of a Chern-Simons term. We will return to this possibility when we consider perturbations breaking $\mathbf{\hat{T}}$ in Sect. \ref{BGapSect}, where we shall see that it plays an important role in the topological nature of the theory.  

In summary, we can understand the exact ground state of (\ref{eqn:Kitaev-honeycomb}) -- a phase whose propagating degrees of freedom consist of Majorana fermions 
coupled to a $Z_2$ gauge field --  as a rather special incarnation of the $Z_2$ spin liquid: a spin-polarized $p$-wave superconductor.
In this description, we arrive at Majorana fermions
not by expressing the spins directly in a Majorana basis,
but rather by starting with Dirac fermions coupled to an $SU(2)$
gauge field and choosing a mean-field solution which breaks the 
gauge symmetry.  The $Z_2$ flux is thus the superconducting vortex,
while the $Z_2$ charge carried by the Majorana fermions reflects the
fact that the superconducting state conserves charge modulo $2$.

\section{T-Breaking Perturbations: The gapped B$^*$ phase}
\label{BGapSect}

In the previous section, we showed that one way to open a gap in the B phase -- by merging the two nodes --  can be understood as a transition between a nodal and nodeless superconductor.  This drives the system into the
A phase. There is, however, a second way to open a gap: we may add another
pairing term to the effective Hamiltonian (\ref{Eq_Hdd}), which will fully gap the spectrum provided that the corresponding gap does not vanish
at the Dirac points.  Here we focus on this latter gapped phase,
and discuss its topological properties.

As noted in Sect. \ref{SymSect}, this second gapped phase necessarily breaks one of the two discrete symmetries of the mean-field solution -- and hence the physical time-reversal symmetry of the spin model -- since we must include couplings between sites on the same sublattice.  Here we will focus on the case of broken $\mathbf{\hat{S}}$, as this can be realized by adding a $3$-spin interaction which commutes with the Hamiltonian (\ref{eqn:Kitaev-honeycomb}).

\subsection{Mean-field theory with $\mathbf{\hat{T}}$ -breaking terms} 

\begin{center}
\begin{figure}[h!]
\includegraphics[height=2in]{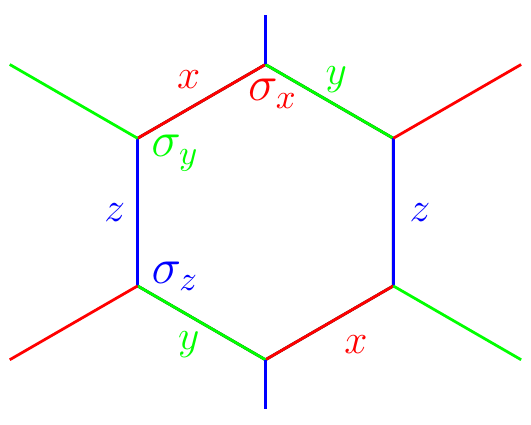}
\caption{ \label{ComFig2} [Color Online]  The $3$-spin interaction which breaks $\mathbf{\hat{T}}$ but commutes with the conserved $6$-spin product on each plaquette (see Fig. \ref{ComFig})\cite{Joost}.   The operator is constructed by taking the product of spin operators on three adjacent vertices, where the direction of the central spin is that associated with the `external' edge at the vertex, while the two external spins match the edges joining their associated vertices to the central vertex.  In the figure the central vertex is at the upper left, and the operator is the product of $\sigma_z$ at the lower left vertex, $\sigma_y$ at the upper left vertex, and $\sigma_x$ at the top vertex.     }
\end{figure}
\end{center}

In terms of the original spin degrees of freedom, the $\mathbf{\hat{T}}$ -breaking term  
we must add to enter the B$^*$ phase is:
\be \label{Eq_SpinAdd}
\frac{J'}{2}  \left( \sum_{ \vec{r}_{ik} =  \hat{x}  } S^x_i S^z_j S^y_k+ \sum_{ \vec{r}_{ik} = \hat{l}_1  } S^z_i S^y_j S^x_k  + 
\sum_{ \vec{r}_{ik} = \hat{l}_2  } S^z_i S^x_j S^y_k \right )  
\ee 
(see Figure \ref{ComFig2}).
It is easy to see that this commutes with the plaquette product of spins (\ref{Eq_Pcons})\cite{Joost}, and hence preserves the $Z_2$ vorticity on each plaquette.  Hence it also commutes with the full Hamiltonian-- though not individually with the spin bilinears on each edge.  

Expressing the spins in terms of Dirac fermions yields a $6$-fermion interaction.  Though we cannot perform the analogue of an exact Hubbard-Stratonovich transformation for the resulting action, which contains  both $4$ and $6$ fermion terms, at small $J'$ it is possible to evaluate its effect on the mean-field solution in a controlled way (see Appendix \ref{BMFSect}).  We find that (consistent with the treatment of Ref. \onlinecite{Kitaev06a}) the effect of such a term is to induce second neighbor hopping and superconducting terms, without altering the rest of the band structure (except for an overall rescaling of the bandwidth).  

We therefore begin by studying the resulting mean-field Hamiltonian.  
The $3$-spin interaction introduces the following quadratic fermion terms for the down spin band:
\ba \label{Eq_HMF2}
H^{(1)}_{MF} &=&  \frac{J'}{8} (- \sin q_x + \sin \vec{q} \cdot \hat{l}_1 -  \sin \vec{q} \cdot \hat{l}_2) \left[ - \Psi^\dag_q \mu_z \Psi_q \right . \n
&& \left.+ \Psi^\dag_q   (\Psi^\dag_{-q})^T+ h.c. \right ] \ \ \ .
\ea
where $\Psi$ was defined in Eq. \ref{Eq_PsiSpinor}.  %{Eq_Hdd}.
As shown in Appendix  \ref{BMFSect}, the perturbation (\ref{Eq_SpinAdd}) does not alter the mean-field Hamiltonian of the up spins, which therefore maintain their flat band structure and remain localized on $x$- and $y$-links. In addition, the new couplings do not disrupt the pair of flat spin down bands.  Thus the basic structure of the initial mean-field solution is preserved, and the only effect of the interaction (\ref{Eq_SpinAdd}) at mean-field is to alter the structure of the dispersing spin-down band.

The new effective mean-field Hamiltonian for the spin-down fermions therefore has the form:
\begin{multline}
\label{Eq_Hdd2}
H_{\rm down}  = 
 \Psi^\dag_{q} \left[ \epsilon_q^{(x)} \mu_x+ \epsilon_q^{(y)} \mu_y+ \epsilon_q^{(z)} \mu_z \right ]   \Psi_{q} \\
  +\Psi^\dag_{q}  \left(\Delta^{(s)}_q \mu_y + \Delta^{(t)}_q \mu_x \right ) (\Psi^\dag_{-q})^T + h.c. \\
   + \tilde{\Delta}^{(p)}_q  \Psi^\dag_q   (\Psi^\dag_{-q})^T + \text{h.c.} \\
  +  \frac{J_z}{8} \left (2 - \frac{J}{J_z}  \right )  \Psi^\dag_{q} \mu_y \Psi_q
 \end{multline}
 with $\epsilon_q^{(x,y)}, \Delta_q^{(s,p)}$ given in Eq.s (\ref{DnDis}-\ref{DnSC}), and
 \be
 \epsilon_z = 
 \tilde{\Delta}^{(p)} = \frac{J'}{8} \left( - \sin q_x + 2 \sin \frac{q_x}{2} \cos \frac{ \sqrt{3} q_y}{2} \right )   \ \ \ .
 \ee
In the vicinity of the Dirac cone, for $J_{x,y,z} \equiv J$, this gives:
\begin{multline}
\label{Eq_HddEff2}
H_{\rm down}  = 
-  \Psi^\dag_{q} \left[  \frac{\sqrt{3} }{32} J  q_y \mu_x- \frac{\sqrt{3} }{32}J  q_x \mu_y+  \frac{J}{16} \mu_y \right.\\
\left. + \left (\frac{3 \sqrt{3}}{64} J' q^2 - \frac{3 \sqrt{3}}{16} J' \right)\mu_z  \right ] \Psi_{q} \\
+ \frac{J\sqrt{3}}{32}   \Psi^\dag_p   \left[ p_y 
   \mu_x - p_x \mu_y \right ] (\Psi^\dag_{-p})^T + \text{h.c.} \\
   - \frac{J}{16}  \Psi^\dag_p  \mu_y (\Psi^\dag_{-p})^T + \text{h.c.}  \\
+    ({3/8 \sqrt{3} J'^2}q^2 + 3 \sqrt{3}J'/2) \Psi^\dag_q   (\Psi^\dag_{-q})^T + h.c.
 \end{multline}
which we can view as a mixed $s$- and chiral $p$-wave superconductor.
This term opens a gap at the Dirac cone, so that the system
is now fully gapped. We discuss the consequences in the next subsection.

\subsection{Topological features of the gapped B phase}
\label{Sect_TopoGauge}

Thus far, we have established that adding the spin interaction (\ref{Eq_SpinAdd}) has the effect, at mean-field, of breaking $\mathbf{\hat{S}}$ and opening a gap in the spectrum of the dispersing Majorana mode ($c$), whilst leaving the band structure of the localized Majorana modes ($b^{x,y,z}$) unchanged.  We will now see how this perturbation leads to a topological phase with $0$-energy Majorana fermions bound to vortices, exactly as in the spinless $p + ip$ superconductor of Read and Green\cite{ReadGreen}.  

%Though the topological nature of the gapped B phase was elucidated by Kitaev\cite{Kitaev06a}, the fermionization we use here permits us to argue the existence of these Majorana bound states in two interesting ways. First, we will show how, in the presence of vortices,  the mean-field model can be viewed as a network of $1$D Majorana wires joined at trivalent vertices -- exactly as described in a recent proposal\cite{Alicea,DasSarma} for realizing topological Majorana fermions in arrays of $1$D superconducting wires.  Second, we will derive an effective Chern-Simons action for the gapped phase, giving a gauge-theoretic picture of the bound Majorana modes.    These two perspectives clarify the relationship between the spin model discussed here and other methods for obtaining phases with the same topological order -- which is the primary objective of this section.  

The simplest way to identify the nature of the B$^*$ phase is to consider
the Hamiltonian (\ref{Eq_H1Band}), where the B phase
is a $p_y$ superconductor. The perturbation modifies the Hamiltonian
according to:
\begin{multline}
\Delta_q \rightarrow \Delta_q - i\frac{J'}{4} \sin \frac{q_x}{2}  \left( \cos \frac{\sqrt{3} }{2} q_y  - \cos \frac{q_x}{2} \right )\\
\approx - i \,\text{sgn}({q_x})\frac{J'}{4} \left(1+\mbox{$\frac{J_z}{2J}$}\right)
\sqrt{1-\left(\mbox{$\frac{J_z}{2J}$}\right)^2}
\end{multline}
In the second line, we have approximated $\Delta_q$ by its value in the
vicinity of the nodes. From this expression, we see that
this is an $ip_x$ superconducting gap which opens
up a gap at the nodes. 

As noted previously, in the nodal B phase, the `chemical potential'
$\mu = 2J-J_z$ lies in the band. Thus, when the gap is opened,
the system goes into the `weak-coupling' $p+ip$ superconducting
phase. As $\vec{q}$ ranges over the Brillouin zone,
the vector $(\text{Re}\Delta_q,\text{Im}\Delta_q,\epsilon_q)/
(\epsilon_q^2 +|\Delta_q|^2)^{1/2}$
wraps around the sphere. The corresponding winding number
cannot be changed without closing the gap, i.e. without going
through a phase transition.

Conversely, when the $3$-spin interaction is included in the A phase,
the chemical potential lies below the band. For sufficiently small $J'$,
$(\text{Re}\Delta_q,\text{Im}\Delta_q,\epsilon_q)/
(\epsilon_q^2 +|\Delta_q|^2)^{1/2}$ remains in the northern
hemisphere, and thus has winding number zero.
Thus, this is the strong-pairing phase of
the chiral $p$-wave superconductor. In other words,
including a weak $\mathbf{\hat{S}}$-breaking
perturbation in the A phase leaves the system in the
A phase.
 
Once we have identified the B$^*$ phase with the weak-pairing phase of the chiral $p$-wave superconductor, we are faced with the following riddle: in its usual incarnation, the superconducting coherence length is assumed to be much larger than the lattice scale, so that vortices are well-modeled by a continuum theory.  In particular, the vortex will have a core which is in the normal state.  The argument put forth by Read and Green\cite{ReadGreen} to show that in the weak-pairing phase a $0$-energy Majorana fermion is bound to the vortex core relies on the existence of a `domain wall' between the vortex core and the superconductor in an essential way.  Since phase B$^*$ is known to have the same topological order as the chiral $p$-wave superconductor, in which the existence of Ising anyons is due to the fact that these $0$-energy Majorana fermions are bound to the vortex cores, we expect a similar phenomenon.  In the lattice model at hand, however, a vortex exists on a single plaquette, and there is no vortex core.  How, then, do the Majorana fermions become bound to these vortices?

%\begin{center}
%\begin{figure}[h]
%\includegraphics[height=1.25in]{Vortex.pdf}
%\caption{ \label{Fig_Vortex} Creating $Z_2$ vortices.  Thickened bonds indicate links on which the sign of $t$ and $\Delta$ has been flipped relative to the mean-field ground state.  A plaquette bordered by an odd number of dark bonds contains a $Z_2$ vortex (shown here as red circles), since a fermion hopping around this plaquette incurs a Berry phase of $\pi$.  Each vortex is associated with a chain of darkened bonds, which may end at the boundary of the system or on another $Z_2$ vortex.  }
%\end{figure}
%\end{center}

One answer to this question comes from studying the long-wavelength gauge theory.
First, we observe that the key effect of the $\mathbf{\hat{T}}$-breaking $3$-spin interaction is that it induces a mass term $m_{( \frac{4 \pi}{3}, 0)} = - m_{( \frac{-4 \pi}{3}, 0)} = \frac{ 3 \sqrt{3} }{2} J'$ at the two nodes in the Brillouin zone.
As discussed perviously, the low-energy effective theory is that of a single
species of massive Dirac fermion.  
If we integrate it out, then as shown explicitly in Appendix \ref{FermiLoopSect},
the $1$-loop effective action for the gauge fields is precisely what we
would expect from a single Dirac cone, except that, since U(1)
is broken down to $Z_2$, a Higgs mass is also generated:
\begin{multline}
\mac{L}_{g}^{(1 \mbox{ loop} )} = 
\frac{1}{2} |\Phi |^2 \,A_\mu A^\mu -
\frac{1}{ 4 \pi m} F^{ \mu \nu} F_{\mu \nu}\\ + \frac{m}{|m| } \frac{1}{ 8\pi} \epsilon_{\mu \nu \lambda} A^\mu\partial^\nu A^\lambda \ \ \ .
\end{multline}
In other words, we obtain the usual Higgs mass term, the
field-strength tensor squared, and a Chern-Simons term with
level $\frac{1}{2}$ (as usual from a single Dirac cone\cite{WWZ}).
The Higgs mass is proportional
to the condensate fraction $ |\Phi |^2$, and is crucial outside a vortex. However, in a vortex
core, the condensate vanishes. We will assume that the Higgs
mass can be neglected in the core.
Thus, in a vortex core, we have
\be
\frac{ \delta \mac{L} } {\delta A^\mu} =
\frac{m}{|m| } \frac{1}{ 8\pi} \epsilon_{\mu \nu \lambda}
\partial^\nu A^\lambda + J_\mu
\ee
where $J_\mu$ is the fermion current,
and we have used $\partial^{\nu} F_{\mu \nu} =0$.  
Taking $\mu =0, m>0$, we obtain the constraint:
\be  \label{Eq_CSConstr}
\frac{1}{4 \pi} \mathbf{B}^z_{\vec{R}} = \rho_{\vec{R}} 
\ee
where $\rho \equiv J_0$.  In the case at hand, we have
\be 
\rho_q = \sum_{i=1,2} \sum_{k} \left [  f^\dag_{k, i} f_{ k-q,i} +  f_{-k, i} f^\dag_{ -k+q,i} \right ] 
\ee
(Here $k$ is technically restricted to momenta near the Dirac cone; more generally, we sum over only half the Brillouin zone.) The rather counter-intuitive fact that {\it holes} at the left Dirac cone carry the same charge as particles at the right Dirac cone results from the fact that the two cones have opposite chirality.  

The density $\rho_q$ is a sum of the density of particles at the right Dirac cone, and holes at the left Dirac cone.  The creation operator associated with this density is the Majorana fermion $c_q = i \left(  f^\dag_{q i} - f_{-qi}   \right)$, which simultaneously creates a particle at $q$ and a hole at $-q$.  
%\be
%\rho_q  \left ( f^\dag_{q i} - f_{-qi}   \right)
%= \left ( f^\dag_{q i} - f_{-qi}   \right) \left (\rho_q  -1  \right) \ \ \ .
%\ee
Hence Eq. (\ref{Eq_CSConstr}) tells us that there is a Majorana fermion $c$ bound to every half-flux quantum.  These half-flux quanta are precisely the $Z_2$ vortices of the superconductor; hence we conclude that there is a Majorana fermion $c$ bound to each $Z_2$ vortex.

\section{Spin-Density Wave States}
\label{sec:SDW}

As described in the previous section, a $T$-breaking
$3$-spin term opens up a gap which can be written
as follows in terms of the $\tilde{\chi}$ fermions,
\begin{equation}
\tilde{\chi}_p =
\begin{pmatrix}
\eta^{}_{\vec{Q}/2+\vec{p}} \cr \eta^\dagger_{-\vec{Q}/2-\vec{p}}
 \end{pmatrix}
.
\end{equation}
At the isotropic point, $J_x = J_y = J_z$, $\vec{Q}/2 = (\frac{4 \pi}{3}, 0)$.
The Hamiltonian in the B phase can be written in the form
\be
H =  \sum_{\vec{p}}
\tilde{\chi}^\dag_p \left [  v p_y \tau_y +  v p_x  \tau_z
\right ] \tilde{\chi}_p    
\ee
where $v=\frac{ \sqrt{3} J}{2} $ at the isotropic point.
The Dirac mass term generated by the three-spin interaction
is of the form:
\be
H_{\rm D.M.} =  m \sum_{\vec{p}} \tilde{\chi}^\dag_p \tau_y \tilde{\chi}_p.
\ee
where $m=3J'/2$.

However, this is not the only possible term which can
open a gap at the nodes of the B phase. The other possible term
is ($W$ is a coupling which we introduce to parametrize the
strength of this term):
\ba
H_{\rm pair} &=&   W \sum_{\vec{p}} \tilde{\chi}^T_p i\tau_y \tilde{\chi}_{-p}
+ \text{h.c.} \cr
&=& 2W \sum_{\vec{p}} \eta^\dagger_{-\vec{Q}/2+\vec{p}}  \eta_{\vec{Q}/2+\vec{p}}
+ \text{h.c.}\cr
&=& 4W \sum_{\vec{p}} \left[c_{\vec{Q}/2-\vec{p}, 1} c_{\vec{Q}/2+\vec{p}, 1}
+ (1\rightarrow 2)\right. \cr
& & {\hskip 1 cm}\left. + ic_{\vec{Q}/2-\vec{p}, 1} c_{\vec{Q}/2+\vec{p}, 2}
+ (1\leftrightarrow 2)\right] + \text{h.c.}\cr
&=& -4W \sum_{\vec{p}}\Bigl[ f^\dagger_{\vec{Q}/2-\vec{p}, 1}
f^\dagger_{\vec{Q}/2+\vec{p}, 1} - f^\dagger_{\vec{Q}/2-\vec{p}, 1}
f_{-\vec{Q}/2-\vec{p}, 1} \cr
& &   {\hskip 2 cm} + \cdots \Bigr] + \text{h.c.}
\ea
Thus, such a mass term breaks translational symmetry.
It includes terms which induce superconductivity at non-zero wavevector
as well as terms which induce a spin-density wave at wavevector
$\vec{Q}$. We can imagine that a spin-spin interaction which
is added to the Kitaev model as a perturbation will, upon
decoupling, generate such a mass term.
However, since the density-of-states at the nodes is zero,
interactions will only generate such a term at O(1) coupling strength
(not at infinitesimal coupling, as would the case for a Fermi
surface instability). At O(1) coupling strength, there is no reason to
focus on the the nodal regions, so many other instabilities could
also occur. It is possible that, in a large-$N$ version of this model,
such a translational-symmetry-breaking instability will occur
at weak-coupling.

Similar but distinct spin-density-wave states have recently been discussed
in the context of a hybrid Kitaev-Heisenberg model in 
Refs. \onlinecite{Chaloupka10,Jiang11}.

\section{Discussion}

In describing the spin-liquid ground states of the various phases of Kitaev's honeycomb model using the slave-fermion approach, we may learn several things about the nature of the phases of this model, their potential stability to perturbations away from the solvable point, and their precise relationship to other phases of matter which exhibit similar physics.  

First, the fermionic mean-field theory allows us to relate the various phases of the Kitaev model to the ground states of different Bogoliubov-de-Gennes Hamiltonians.  This can be done in two different ways:
(1) in terms of the fermions $f_{\uparrow,\downarrow}$
introduced in Eq. \ref{Eq_Spins} and (2) in terms of
the fermions $\eta$ introduced in Eq. \ref{eqn:eta-def}.
The latter are formed from the propagating part of
$f_{\downarrow}$. Each way has its conceptual and technical
advantages, as we have seen.

\begin{center}
\begin{figure}[h!]
\includegraphics[height=2.5in]{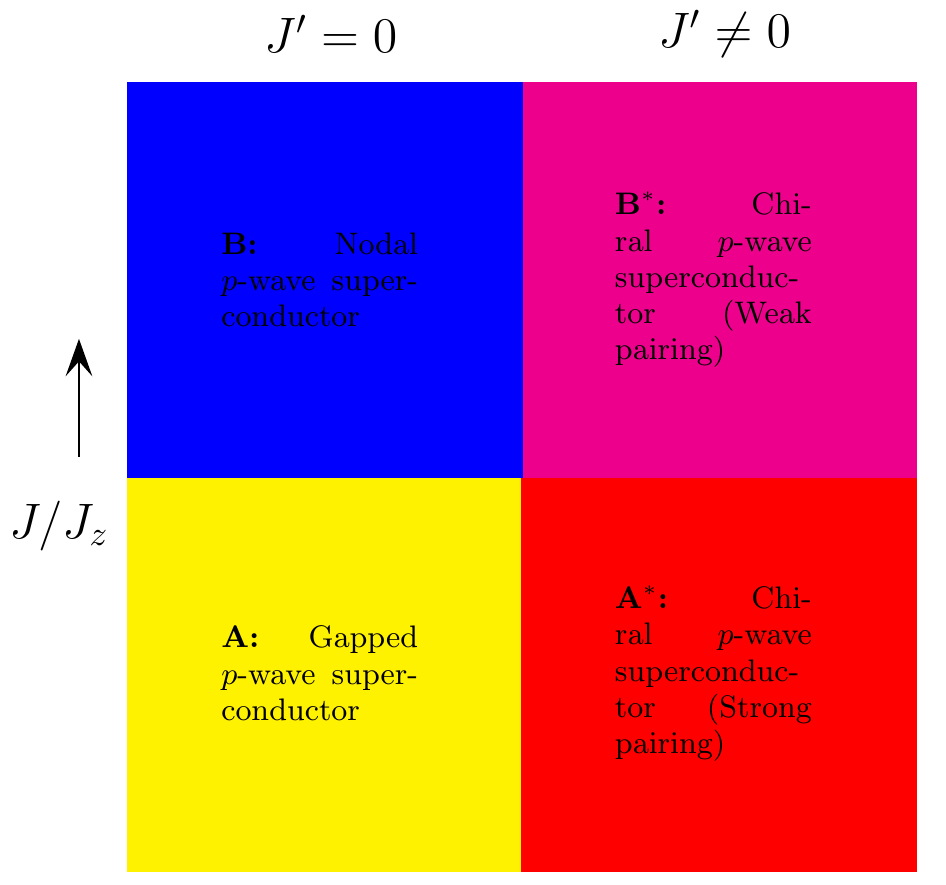} 
\caption{ \label{Fig_Phase} Schematic phase diagram of the Kitaev honeycomb model, and the corresponding superconducting phases.  The phase is determined by the ratio $J/J_z$, and by whether the coefficient $J'$ of the $3$-spin interaction is non-vanishing.  }
\end{figure}
\end{center}

The mean field phase diagram is summarized
in Figure \ref{Fig_Phase}, which can be interpreted in terms
of the $\eta$ fermions as follows. The A phase, in which the nodes of the
superconductor do not intersect the Fermi surface, is adiabatically connected to an $s$-wave superconductor.  The B phase is a nodal $p$-wave superconductor.  The B$^*$ phase is the weak-pairing phase of a chiral $p$-wave superconductor, with the consequent Ising topological order.  The A$^*$ phase is the corresponding strong-pairing chiral $p$-wave superconductor phase. As a result of the strong-pairing
nature of this phase, the topological order is, in fact, again that of an $s$-wave
superconductor. The reason for this is that, at the mean-field level (i.e. when treated
as a free fermion problem) the A and A$^*$ phases can be adiabatically
deformed into each other, so the line between them in Figure \ref{Fig_Phase}
is a crossover line. On the other hand, the other transitions in
Figure \ref{Fig_Phase} are genuine phase boundaries which
are essentially the same as the corresponding transitions in the
superconductor. One important difference needs to be emphasized.
In a two-dimensional superconductor with a three-dimensional
electromagnetic field, there is a gapless plasmon. Thus, a thin
superconducting film is not fully gapped, even though its fermionic spectrum
is fully-gapped. However, in the Kitaev honeycomb lattice model,
the gauge field is two-dimensional. Consequently, the plasmon is
gapped and the system is fully-gapped.

Although the SU(2) mean-field theory
described here is clearly more complicated than
Kitaev's at the soluble point, it has the
salient virtue that it is well-suited to perturbing away
from the soluble point -- partiaularly in the gapless B phase.

It is interesting to consider the fate of the phase diagram shown in
Figure \ref{Fig_Phase} when the spin Hamiltonian is deformed away from
the exactly solvable point. The fully-gapped phases -- the A and
B$^*$ phases -- will be robust against small perturbations by virtue
of their energy gaps. So long as the gauge symmetry is broken to $Z_2$ by the saddle point solution, the gauge field is gapped, and
we do not expect fluctuations to lead to confinement.
Therefore, the model still admits effective spinon excitations,
and the topological order of the spin liquid will be robust to gauge field fluctuations. Since the spinons are also gapped in these phases, they
are stable against adding weak interactions between the fermions.
The gapless B phase is a little trickier. Since
the gauge field is fully gapped, we believe that the gauge field
action is robust against small perturbations. The fermions, on the other hand,
are gapless. However, since they have a single gapless Dirac point (rather
than a Fermi surface), weak interactions between the fermions are
irrelevant by power-counting. This is the reason that
an $SU(2)$-invariant Heisenberg perturbation does not
lead to a phase transition until the perturbation is sufficiently strong.
Thus, we could say that the stability of the gapless B phase
relies on phase space limitations.
However, as we have seen, although the gapless B phase
is stable against weak perturbations, some features of
the soluble point are not generic to this phase.
For instance, a magnetic field will make the
spin-spin correlation functions have a power law,
rather than short-ranged, form.

The fact that the bosonic fluctuations are all gapped does not, however, prevent the theory from acquiring a new lowest-energy saddle point if we deform
far enough away from the solvable model.
For instance, as we have discussed, the gapless B phase
can acquire a gap by an alternative method: the development of
a spin-density wave, as discussed in Section \ref{sec:SDW}.
Various perturbations of the Kitaev model,
including a Heisenberg interaction \cite{Chaloupka10,Jiang11}
can lead to such an instability. Furthermore, it is well known\cite{Dimers,AffleckMarston} that symmetric spin-liquid states are often
prone to dimerization instabilities, in which the spins pair with neighbors in a valence-bond crystal which breaks a lattice symmetry.  
Away from the solvable limit, therefore, it is likely that the phase diagram will also include some such valence-bond crystal states.  At the symmetric point, the model has a spin-orbit type $3$-fold rotation symmetry (entailing a $3$-fold lattice rotation about a vertex, coupled with a global spin rotation of the form (\ref{Eq_Residuals})) which makes the saddle point perturbatively stable -- though in principle lower-energy symmetry-breaking saddle points might exist.  
 Away from the isotropic point $J_x = J_y = J_z$, such states need not break any symmetries of the Hamiltonian, so that symmetry does not prevent the saddle point from flowing to such a valence-bond crystal upon including fluctuations of the amplitudes of the mean-field hopping and superconducting terms.

The fact that the exact ground state of (\ref{eqn:Kitaev-honeycomb}) can be correctly described in the slave-fermion mean-field approach used here is also interesting in its own right.  As discussed above, since the mean-field state is a Higgs phase of the gauge field, the model is in a r{\'e}gime where the spin-liquid saddle point is most likely to be stable.  
Even in this case, however, examples of Hamiltonians where the exact ground state can be shown to be a spin liquid are rare.  The Kitaev model is thus a potential testing ground for the slave-fermion approach, since we may begin with a Hamiltonian for which it is demonstrably valid, and consider the fate of the ground state under various perturbations.  In particular, on general grounds\cite{Hastings} we expect that for small perturbations which do not close the gap in the spectrum, the slave-fermion mean-field theory will continue to capture the topological order of the gapped phases.

Another interesting prediction of the slave-fermion approach is that near the solvable point, the Kitaev model becomes a  superconductor upon doping.  Specifically, we imagine starting with a Mott insulator whose effective Hamiltonian at half-filling is given by (\ref{eqn:Kitaev-honeycomb}).  After doping away from half-filling we must account for the fermion hopping terms, leading to a $t-J$ model, with the spin Hamiltonian given by (\ref{eqn:Kitaev-honeycomb}).  Following the prescription used to study the cuprates\cite{LeeNagaosa}, we may decompose the spin operators as in Eq. (\ref{Eq_Spins}), and express the electron operator as
\be
c^\dag_{i \sigma }= f^\dag_{i \sigma} b_{i \sigma}
\ee
with the constraint
\be
f^\dag_{i \uparrow}f_{i \uparrow} + f^\dag_{i \downarrow}f_{i \downarrow} + b^\dag_i b_i = 1 \ \ \ .
\ee
It follows that, at temperatures below the Bose condensation temperature of the bosons, and at sufficiently low dopings that the mean-field solution described above is a good approximation for the spinons ($f_{i \sigma}$), the superconducting order parameter is:
\ba
\Delta^{phys}_{k; \sigma, \sigma'} &=& \langle c^\dag_{k \sigma} c^\dag_{-k \sigma'} \rangle =  \langle f^\dag_{k+q,  \sigma} f^\dag_{-k-q, \sigma'} \rangle  \langle b_{-q \sigma} b_{q \sigma'} \rangle \n
&=& \Delta_{k; \sigma, \sigma'} \rho_s
\ea
where $\rho_s$ is the bosonic superfluid density.  Thus the momentum dependence of the physical superconducting order parameter is set by that of the mean-field superconducting order parameter $\Delta$ for the fermionic spinons $f$.  For the Hamiltonian (\ref{eqn:Kitaev-honeycomb}), this predicts spin-triplet superconductivity
(with equal spin pairing), with a mixed singlet and triplet pseudospin order parameter.  

Finally, it is interesting to compare the mean-field ground state of
the Kitaev model with existing proposals for generating the B$^*$ phase's
topological Majorana fermions in physical materials.
The mean-field Hamiltonian of the B phase is manifestly equivalent to
a $p+ip$ superconducting state of spin-polarized fermions \cite{ReadGreen}.
It also has an interesting relation to the the effective Hamiltonian of Fu and Kane\cite{FuKane} for surface states of a topological insulator in the presence of induced $s$-wave superconductivity. In the absence of superconductivity,
these surface states form a single Dirac fermion. This Dirac
fermion is analogous to the Dirac fermion which we have in
the gapless B phase. If a magnetic film is brought into contact
with the topological insulator, and the magnetic moment is perpendicular
to the interface, then the resulting term in the Hamiltonian is
a Dirac mass term, which breaks time-reversal symmetry
and opens a gap. This is analogous to the $3$-spin term in the
Kitaev model, which opens a gap and drives the system into the
B$^*$ phase. Note that this term in the Kitaev model is {\it not}
analogous to the term generated by an $s$-wave superconducting
film on the surface of a topological insulator. Instead, $s$-wave
superconductivity on the surface of a topological insulator is analogous
to a term ${\tilde \chi}^T_p {i\tau_y} {\tilde \chi}_p + \text{h.c.}$,
which is a down-spin density wave at wavevector $(8\pi/3,0)$
at the symmetric point ${J_x}={J_y}={J_z}$.

In all cases, the essential ingredients for generating topological Majorana fermions are a $2$-band model in which the band structure is that of a massive Dirac fermion, and with induced superconductivity.  As we described in Sect. \ref{Sect_TopoGauge} above, the massive Dirac fermion in all of these models is implicitly coupled to a gauge field, since it forms a superconducting state.  The fermion mass therefore generates a Chern-Simons term in the effective gauge-field action, which has the effect of binding a half-quantum vortex to each charge, since there is only a single Dirac cone.  The charge which is bound in the superconducting state is a Bogoliubov-de-Gennes quasi-particle, rather than a fermion -- which, when the superconducting order parameter has a $p$-wave component, binds a Majorana fermion to the vortex.  

\acknowledgements
We gratefully acknowledge the hospitality of the Aspen Center
for Physics, where part of this work was completed.
We thank John Chalker, John Cardy, and Simon Trebst for discussions.
C.N. has been supported in part by the DARPA QuEST program.

\appendix

\section{Mapping between $SU(2)$ and Majorana fermionizations}
\label{sec:Relation-to-Majorana}

Here we explain in more detail the correspondence between the fermionization (\ref{Eq_Spins}) and the Majorana fermionization employed by Kitaev\cite{Kitaev06a}.  We begin with the mean-field correspondence:
\ba \label{Eq_Mar2Dir2}
b^x_{{\bf q}i} =  i \left( f^\dag_{{\bf q} i \uparrow } - f_{-{\bf q} i 
\uparrow } \right )&\ \ \ \ \
b^y_{{\bf q} i} = f^\dag_{{\bf q} i \uparrow } + f_{-{\bf q} i \uparrow}    \n
 b^z_{{\bf q} i} =  f^\dag_{{\bf q} i \downarrow} + f_{-{\bf q} i \downarrow } & \ \ \ \ \ c_{{\bf q} i} =   i \left( f^\dag_{{\bf q} i \downarrow } - f_{-{\bf q} i \downarrow } \right ) \ \ \ . 
 \ea
which gives a mapping between unprojected spinful fermions and unprojected Majorana fermions.  This mapping is not unique, as each Majorana fermion can be represented by any linear combination:
\be
c_{{\bf q}i} = f^\dag_{{\bf q} i \sigma } e^{i \phi} + h.c.
\ee
and any choice of $4$ such combinations which mutually anti-commute could be associated with $\{ b^{x}, b^{y},b^{z}, c \} $.  However, this difference is not physical, as all such mappings are equivalent under $SU(2)$ gauge transformations.

The mapping (\ref{Eq_Mar2Dir2}) does not preserve the form of the unprojected spin operators, however.  
Specifically, the fermionization (\ref{Eq_Spins}) gives
\ba
S^x_i & =& \left( f^\dag_{i \uparrow} f_{i \downarrow} + f^\dag_{i \downarrow} f_{i \uparrow} \right ) \n
S^y_i & =& - i \left( f^\dag_{i \uparrow} f_{i \downarrow} + f^\dag_{i \downarrow} f_{i \uparrow} \right ) \n
S^z_i & =& \left( f^\dag_{i \uparrow} f_{i \uparrow} - f^\dag_{i \downarrow} f_{i \downarrow} \right ) \n
\ea
while Kitaev's Majorana fermionization stipulates:
\ba \label{TheSpins}
\tilde{S}^x_i  =& i b^x_i c_i &= - i   \left( f^\dag_{ i \uparrow } - f_{ i \uparrow } \right ) 
\left( f^\dag_{ i \downarrow } - f_{ i \downarrow } \right ) \n
\tilde{S}^y_i  =& i b^y_i  c_i & =  - \left( f^\dag_{ i \uparrow } + f_{ i \uparrow } \right ) \left( f^\dag_{ i \downarrow } - f_{ i \downarrow } \right )  \n
\tilde{S}^z_i  =&i b^z_i c_i & =   - \left( f^\dag_{ i \downarrow} + f_{i \downarrow } \right )  \left( f^\dag_{ i \downarrow } - f_{ i \downarrow } \right )  \n
\ea
This gives:
\ba
\tilde{S}^x_i &=& -S^y_i -i  \left( f^\dag_{i \uparrow} f^\dag_{i \downarrow} + f_{i \uparrow} f_{i \downarrow} \right)\n
\tilde{S}^y_i &=& S^x_i   - \left( f^\dag_{i \uparrow} f^\dag_{i \downarrow} -  f_{i \uparrow} f_{i \downarrow} \right)   \n
\tilde{S}^z_i &=& -S^z_i  +\left (n_{i \uparrow} + n_{i \downarrow} -1 \right )   
\ea
which, after a gauge transformation to rotate the spins and eliminate the extra phases, differs from the spin operators (\ref{TheSpins}) by terms which vanish under projection onto the physical Hilbert space.  
It is these extra terms which lead to the fact that the mean-field Hamiltonian (\ref{Eq_MarH}) does not conserve 
$b^x_i b^x_j$ on $x$-links (and similarly for $y$ and $z$) so that it is not obvious that the mean-field theory captures the essentials of the spin-spin correlations, as it is in the Majorana description.

However, one way to view the equivalence of the two descriptions is via the wave functions that they produce {\it after} projection.  The Majorana projector is:
\ba \label{Eq_KitCon}
D_i& \equiv& b_i^x b_i^y b_i^z c_i = \hat{\mathbf{1}} \ \ \ .\\
& =&- \left( f^\dag_{i  \uparrow } + f_{ i \uparrow}  \right )  \left( f^\dag_{ i \uparrow } - f_{ i 
\uparrow } \right ) \left( f^\dag_{ i \downarrow} + f_{i \downarrow }  \right )  \left( f^\dag_{ i \downarrow } - f_{ i \downarrow } \right )   \nonumber 
\ea
Expanding the constraint in terms of Dirac fermion operators, we obtain
\ba
D_i &=& -( 2 n_{i \uparrow} -1 )( 2 n_{i \downarrow} -1) \n
% &=& -(4  n_{i \uparrow} n_{i \downarrow} - 2 (n_{i \uparrow}+ n_{i \downarrow} ) +1) \n
&=& -2 (n_{i \uparrow}+ n_{i \downarrow}  - 1)^2 +1
\ea
Hence imposing the diagonal $SU(2)$  constraint
\be
n_{i \uparrow}+ n_{i \downarrow}  - 1 =0
\ee
automatically imposes the Majorana constraint $D_i =\hat{\mathbf{1}} $.  

Therefore, if we begin with a mean-field wave-function expressed in terms of the spinful fermions, and project onto the physical Hilbert space of singly occupied states, this is equivalent to studying the same mean-field wave function expressed in terms of Majorana fermions, and applying the projector (\ref{Eq_KitCon}) at each site.  This gives an alternative perspective on why the mean-field theory is exact.

\section{Mean-field theory of the quadratic spin model} \label{AMFSect}

Here we will review the detailed derivation of the mean-field Hamiltonian (\ref{MFHam}).  We will first show how to derive the full effective action, and then present the self-consistent mean-field solution.  

 \subsection{Hubbard-Stratonovich decoupling of the Kitaev model}

In the Dirac fermion basis, the $3$ different types of terms in the Hamiltonian (\ref{eqn:Kitaev-honeycomb}) are:
\begin{widetext}
\ba  \label{Eq_HeisenbergTerms}
\hat{S}^x_i \hat{S}^x_j &=&-\frac{1}{4} \left[ f^\dag_{i \uparrow} f^\dag_{j \uparrow} f_{i \downarrow} f_{j 
\downarrow} +  f^\dag_{i \downarrow}  f^\dag_{j \downarrow} f_{i \uparrow}f_{j \uparrow}
+  f^\dag_{i \uparrow} f_{j \uparrow} f^\dag_{j \downarrow} f_{i \downarrow}
+  f^\dag_{i \downarrow} f_{j \downarrow} f^\dag_{j \uparrow} f_{i \uparrow} \right ] \n
\hat{S}^y_i \hat{S}^y_j &=&-\frac{1}{4}\left[ -f^\dag_{i \uparrow} f^\dag_{j \uparrow} f_{i \downarrow} f_{j 
\downarrow} -  f^\dag_{i \downarrow}  f^\dag_{j \downarrow} f_{i \uparrow}f_{j \uparrow}
+  f^\dag_{i \uparrow} f_{j \uparrow} f^\dag_{j \downarrow} f_{i \downarrow}
+  f^\dag_{i \downarrow} f_{j \downarrow} f^\dag_{j \uparrow} f_{i \uparrow} \right ] \n
\hat{S}^z_i \hat{S}^z_j& =&-\frac{1}{4}\left[  f^\dag_{i \uparrow} f^\dag_{j \uparrow} f_{j \uparrow} f_{i 
\uparrow}
 +f^\dag_{i \downarrow} f^\dag_{j \downarrow} f_{j \downarrow} f_{i \downarrow}
+  f^\dag_{i \uparrow} f_{j \uparrow} f^\dag_{j \uparrow} f_{i \uparrow}
+  f^\dag_{i \downarrow} f_{j \downarrow} f^\dag_{j \downarrow} f_{i \downarrow} \right ]
\ea
where we have used $n_{i \uparrow } = 1- n_{i \downarrow}$ in the last expression.

To decouple the $4$-fermi interactions using Hubbard-Stratonovich fields, we take the Lagrangian:
\ba \label{Eq_HSDec}
\mac{L}_x &=& -\frac{8 (|\Phi_1|^2 + | \Phi_2|^2)}{J_x} +  \Phi_1\left ( f^\dag_{i \uparrow} f_{j \uparrow} + 
f^
\dag_{i \downarrow} f_{j \downarrow} \right) + i \Phi_2 \left ( f^\dag_{i \uparrow} f_{j \uparrow} - f^\dag_{i 
\downarrow }f_{j \downarrow} \right)   + \tilde{ h.c.} \n
&& -\frac{8 (|\Theta_1|^2 + | \Theta_2|^2)}{J_x} +  \Theta_1\left ( f^\dag_{i \uparrow} f^\dag_{j \uparrow} 
+ f^
\dag_{i \downarrow} f^\dag_{j \downarrow} \right) + i \Theta_2 \left ( f^\dag_{i \uparrow} f^\dag_{j 
\uparrow} - 
f^\dag_{i \downarrow }f^\dag_{j \downarrow} \right)   + \tilde{ h.c.} \n
\mac{L}_y &=& -\frac{8(|\Phi_1|^2 + | \Phi_2|^2)}{J_y} +  \Phi_1\left ( f^\dag_{i \uparrow} f_{j\uparrow} + 
f^
\dag_{i\downarrow} f_{j\downarrow} \right) + i \Phi_2 \left ( f^\dag_{i \uparrow} f_{j \uparrow} - f^\dag_{i
\downarrow} f_{j\downarrow} \right)   + \tilde{  h.c. }\n
&& -\frac{8 (|\Theta_1|^2 + | \Theta_2|^2)}{J_y} +  i \Theta_1\left ( f^\dag_{i \uparrow} f^\dag_{j \uparrow} 
+ f^
\dag_{i \downarrow} f^\dag_{j \downarrow} \right) +  \Theta_2 \left ( f^\dag_{i \uparrow} f^\dag_{j 
\uparrow} - f^
\dag_{i \downarrow }f^\dag_{j \downarrow} \right)   - \tilde{  h.c.} \\
\mac{L}_z &=& -\frac{4( |\Phi_1|^2 + | \Phi_2|^2)}{J_z} +  \Phi_1 f^\dag_{i\uparrow} f_{j\uparrow} +  
\Phi_2 f^
\dag_{i\downarrow} f_{j\downarrow} +\tilde{  h.c.}  -\frac{4 (|\Theta_1|^2 + | \Theta_2|^2)}{J_z} +  
\Theta_1 f^\dag_{i \uparrow} f^\dag_{j 
\uparrow} +  \Theta_2  
f^\dag_{i \downarrow }f^\dag_{j \downarrow}   + \tilde{  h.c. } \nonumber
\ea
where the fields $\Phi_i, \Theta_i$ are to be understood as being evaluated on the link in question, and 
the 
$\tilde{ h.c.}$ is the hermitian conjugate with all spin directions reversed. 
We can check that this decoupling gives back the original action by integrating out the bosonic fields.  
For example, completing the 
square for the first line of $\mac{L}_x$ gives:
\ba
\mac{L}_x &=&- \frac{8}{J_x} \left[ \Phi_1-\frac{J_x}{8}\left ( f^\dag_{j \uparrow} f_{i \uparrow} + f^\dag_{j 
\downarrow} f_{i\downarrow} \right) \right ]  \left[ \Phi_1^\dag- \frac{J_x}{8} \left ( f^\dag_{i \uparrow} f_{j 
\uparrow} + f^\dag_{i \downarrow} f_{j \downarrow}  \right) \right ]  
+ \frac{J_x}{8} \left ( f^\dag_{i \uparrow} f_{j \uparrow} + f^\dag_{i \downarrow} f_{j \downarrow}  \right) 
\left 
( f^\dag_{j \uparrow} f_{i \uparrow} + f^\dag_{j \downarrow} f_{i\downarrow} \right) \n
&& - \frac{8}{J_x} \left[ \Phi_2-i \frac{J_x}{8} \left ( f^\dag_{j \uparrow} f_{i \uparrow} - f^\dag_{j 
\downarrow} f_{i
\downarrow} \right) \right ]  \left[ \Phi_2^\dag- i \frac{J_x}{8} \left ( f^\dag_{i \uparrow} f_{j \uparrow} - f^
\dag_{i 
\downarrow} f_{j \downarrow}  \right)\right ] 
-\frac{J_x}{8} \left ( f^\dag_{i \uparrow} f_{j \uparrow} - f^\dag_{i \downarrow} f_{j \downarrow}  \right) \left 
( f^\dag_{j \uparrow} f_{i \uparrow} - f^\dag_{j \downarrow} f_{i\downarrow} \right)    
\ea
\end{widetext}
Integrating out the factors involving $\Phi_1$ and $\Phi_2$ gives a constant; the sum of the remaining 
pieces gives:
\be
\frac{J_x}{4} \left(  f^\dag_{i \uparrow} f_{j \uparrow} f^\dag_{j \downarrow} f_{i\downarrow}  +  f^\dag_{i 
\downarrow} f_{j \downarrow}  f^\dag_{j \uparrow} f_{i \uparrow}  \right)
\ee
as expected.

Now we proceed in the usual way for mean-field theories: namely, the fields $\Theta$ and $\Phi$ have 
gapped amplitude fluctuations, as well as phase fluctuations.  We will thus begin with a mean-field 
solution $\Theta_{\sigma, ij}(t) \equiv 
\Delta_{\sigma, ij}, \Phi_{\sigma, ij}(t) \equiv t_{\sigma, ij}$ which reproduces the quadratic 
fermionic spectrum of the exact solution.
We then consider the fate of the fluctuations of both gapped 
amplitude modes and gapless phase modes about mean-field.

\subsection{Mean-field solution} \label{MFSect}

At mean-field level, the relevant information contained in Eq. (\ref{Eq_HSDec}) is that on each link there 
are 
potentially $4$ bosonic fields: $t_{\uparrow} $ associated with hopping of up spins, $t_{\downarrow}$ 
with 
hopping of down spins (which formally transforms in the opposite way under time reversal), and 
separate 
superconducting order parameters $\Delta_\uparrow, \Delta_{\downarrow}$ for the spin up and spin 
down 
sectors.  Formally, in terms of the fields of the previous section, we take:
\ba
t_{\uparrow} &= \langle \Phi_1 + i \Phi_2 \rangle \mbox{  on $x$ and $y$ links  }  \ \ \ t_{\uparrow} 
&= \langle \Phi_1 \rangle \mbox{  on $z$ links }  \n
t_{\downarrow} &= \langle \Phi_1 - i \Phi_2 \rangle \mbox{  on $x$ and $y$ links  }  \ \ \ t_{\downarrow} 
&= \langle \Phi_2  \rangle \mbox{  on $z$ links }   \n
\Delta_{\uparrow} &= \langle i \Theta_1 + \Theta_2 \rangle \mbox{  on $y$  links  } \ \ \  \Delta_
{\downarrow} &= \langle i \Theta_1 -  
\Theta_2 \rangle \mbox{  on $y$ links } \n
\Delta_{\uparrow} &= \langle \Theta_1 + i \Theta_2 \rangle \mbox{  on $x$  links  }  \ \ \ \Delta_{\uparrow} 
&= \langle \Theta_1\rangle  \mbox{  on $z$ links }  \n
\Delta_{\downarrow} &= \langle \Theta_1 - i \Theta_2 \rangle \mbox{  on $x$ links }  \ \ \ \Delta_
{\downarrow} 
&= \langle \Theta_2\rangle  \mbox{  on $z$ links } 
\ea
 From the Lagrangian (\ref{Eq_HSDec}), the saddle-point equations are:
\ba  \label{Eq_MFs}
 t^{(x,y)}_{\uparrow} &=&  \frac{J_{x,y} }{4}  \langle f^\dag_{j \downarrow} f_{i \downarrow}    \rangle \ \ \ \
  t^{(x,y)}_{\downarrow} =   \frac{J_{x,y,} }{4}  \langle f^\dag_{j \uparrow} f_{i \uparrow}   \rangle  \n
   t^{(z)}_{\uparrow} &=&  \frac{J_{z} }{4}  \langle f^\dag_{j \uparrow} f_{i \uparrow}    \rangle \ \ \ \
  t^{(z)}_{\downarrow} =   \frac{J_{z,} }{4}  \langle f^\dag_{j \downarrow} f_{i \downarrow}   \rangle  \n
\Delta^{(x)}_{\uparrow} &=&  \frac{J_{x} }{4}  \langle f_{j \downarrow} f_{i \downarrow}    \rangle \ \ \ \
  \Delta^{(x)}_{\downarrow} =   \frac{J_{x} }{4}  \langle f_{j \uparrow} f_{i \uparrow}   \rangle  \n
  \Delta^{(y)}_{\uparrow} &=&  \frac{-J_{y} }{4}  \langle f_{j \downarrow} f_{i \downarrow}    \rangle \ \ \ \
  \Delta^{(y)}_{\downarrow} =   \frac{-J_{y} }{4}  \langle f_{j \uparrow} f_{i \uparrow}   \rangle  \n
  \Delta^{(z)}_{\uparrow} &=&  \frac{J_{z} }{4}  \langle f_{j \uparrow} f_{i \uparrow}    \rangle \ \ \ \
  \Delta^{(z)}_{\downarrow} =   \frac{J_{z} }{4}  \langle f_{j \downarrow} f_{i \downarrow}   \rangle   \ \ \ .
\ea

 To satisfy the mean-field conditions (\ref{Eq_MFs}), we take:
\ba \label{Eq_MFUp}
t_{ij, \downarrow} &=& -\Delta_{ij, \downarrow} = \frac{iJ_x}{16}  \ \ \mbox{     on $x$-links } \n
t_{ij, \downarrow}&=&  -\Delta_{ij, \downarrow} = \frac{iJ_y}{16} \ \  \mbox{     on $y$-links } \n
t_{ij, \uparrow} &=& \Delta_{ij, \uparrow} = 0 \ \ \ \  \mbox{     on $z$-links } \n
t_{ij, \uparrow} &=& -\Delta_{ij, \uparrow} = \frac{iJ_x}{16} \ \  \mbox{     on $x$-links } \n
t_{ij, \uparrow}&=&   \Delta_{ij, \uparrow} = \frac{iJ_y}{16} \ \  \mbox{     on $y$-links } \n
t_{ij, \downarrow} &=&  i \frac{J_z}{8}  \ \ \ \ \ \  \Delta_{ij, \downarrow}  = 0 \ \  \mbox{     on $z$-links }
\ea
which gives the mean-field Hamiltonian (\ref{MFHam}).

\subsection{Theory of fluctuations about mean field} \label{FluctApp}

We now turn to the fluctuations about the mean-field solutions.  Since symmetry dictates that these 
cannot change the fermionic band structure, our focus will be to describe the bosonic degrees of
freedom in this theory, and demonstrate that the gauge field is in a Higgsed phase with a residual $Z_2$ symmetry group.  

The Hubbard-Stratonovich decoupling introduces $4$ bosonic fields: $\Phi_{1,2}$, whose saddle-point expectation values are associated with fermion hopping terms; and $ \Theta_{1,2}$,  associated with the spin-triplet superconductivity.  We parametrize their fluctuations according to: 
\ba  \label{Eq_Flucts}
\Phi_{1 ij} &=&  \mp \left( \frac{i}{16} \left( J_x \delta_{ij, x}+  J_y \delta_{ij, y} + 2 J_z \delta_{ij, z} \right )  
e^{i a_{ij }  } + i \phi_{ ij}  \right )\n
\Phi_{2 ij} &=& \pm i \left(  \frac{J_z}{8}\delta_{ij, z}   e^{i \tilde{\theta}_{ij}  } + \tilde{\rho}_{ ij} \right ) \n
\Theta_{1 ij} &=& \pm i  \left(  \frac{J_y}{16}  \delta_{ij, y}    e^{i \theta_{ij }} + \rho_{ ij}  \right )\n
\Theta_{2 ij} &=& \mp i \left( \frac{ J_x }{16} \delta_{ij, x}  e^{i \theta_{ij}} + \rho_{ ij} \right )
\ea 
where the functions $\delta_{ij, x,y,z}$ have support on $x,y$, and $z$ links respectively, and the top 
(bottom) sign is taken for edges 
oriented from sublattice 1(2) to sublattice 2(1).  

The physical interpretation of these fields is as follows.  $\Phi_1$ is associated with the spin-rotation 
invariant hopping terms familiar 
from spinon decompositions of the Heisenberg model\cite{AffleckMarston,WenPSG}.  The  phase 
variables $a_{ij}$ are the spatial 
components of the gauge fields associated with the constraints (\ref{Eq_Constrs}); 
fluctuations in the amplitude of this hopping term are parametrized by the 
scalar $\phi$.  

The remaining terms parametrize fluctuations of a condensed superfluid which breaks the $SU(2)$ 
gauge group down to $Z_2$.   We 
combine the fields associated with $\Theta_1$ and $\Theta_2$, each of which is non-vanishing at 
mean-field either on $ x$ or $y$-
links respectively, into a single pair of scalar fields $\rho, \theta$ defined on all links in the lattice.  Since 
at mean-field, $\Theta$'s 
expectation value generates a spinful superconducting pairing, $\theta$ is the phase of a charged 
superfluid, and hence in the 
condensed phase becomes the longitudinal component of the corresponding gauge field.  $\rho$ 
parametrizes the (gapped) 
fluctuations in this superfluid density.   

That $\Phi_2$, the hopping anti-symmetric in spin, is associated with a charged superfluid is less 
obvious.  We will show shortly, 
however, that $\langle \Phi_2 \rangle$ breaks the off-diagonal generators of $SU(2)$.  As these are not 
the same as the generator 
broken by the superconducting terms, we use a new field $\tilde{\theta}$ to denote the phase 
fluctuations.

To find 
the residual symmetry group, 
we must evaluate the $SU(2)$ flux through each lattice plaquette at mean-field\cite{WenPSG}.  
It is enlightening to express the fermionic degrees of freedom in terms of the usual BCS spinors:
\be
\chi_q = \bv{c} f_{\uparrow, q} \\ f^\dag_{\downarrow, -q} \\ \ev
\ee
which transform under gauge transformations by $e^{i \vec{\alpha} \cdot \vec{\sigma} }$ as
\be
\chi_q \rightarrow e^{i \vec{\alpha} \cdot \vec{\sigma} } \chi_q \ \ \ .
\ee
In this basis, the spin-symmetric and spin-antisymmetric hopping terms can be expressed
\ba
i t_{\uparrow+\downarrow}(ij) \left( f^\dag_{i \uparrow} f_{j \uparrow} +  f^\dag_{i \downarrow} f_{j 
\downarrow} - f^\dag_{j \uparrow} f_{i 
\uparrow} -  f^\dag_{j \downarrow} f_{i \downarrow} \right ) &=&\n
 i t_{\uparrow+\downarrow}(ij) \left( \chi^\dag_i  \chi_j-  \chi^\dag_{j}  \chi_{i} \right ) &&\n
i t_{\uparrow-\downarrow}(ij) \left( f^\dag_{i \uparrow} f_{j \uparrow} -  f^\dag_{i \downarrow} f_{j 
\downarrow} - f^\dag_{j\uparrow} f_{i 
\uparrow} +  f^\dag_{j \downarrow} f_{i \downarrow} \right ) &=& \n
i t_{\uparrow-\downarrow}(ij) \left( \chi^\dag_i\sigma_z \chi_j - \chi^\dag_{j}\sigma_z \chi_{i} \right) &&
\ea
As promised, the first term is gauge invariant under all generators.  The effect of a gauge transformation 
on the second term is to 
conjugate the matrix $\sigma_z$ by $e^{i \vec{\alpha} \cdot \vec{\sigma} }$.  Hence this term is invariant 
under the $U(1)$ subgroup 
comprised of rotations about the $z$ axis, but not under rotations by the two generators $\sigma_x$ 
and $\sigma_y$.  Fluctuations in $
\tilde{\theta}$ are therefore associated with the longitudinal modes of the broken generators $a_{ij}^{(x, 
y)}$.   

The remaining $U(1)$ symmetry is broken by the superconducting terms.  As the pairing occurs here in 
the spin triplet channel, these 
cannot naturally be expressed in the BCS basis; however, they are clearly charged under the residual 
$U(1)$ symmetry $f_{i \sigma} 
\Rightarrow e^{i \alpha_i}f_{i \sigma}$.  Hence the $U(1)$ symmetry is broken to the $Z_2$ subgroup 
$f_{i \sigma} \Rightarrow \pm f_{i 
\sigma}$, which is the residual gauge symmetry of the Hamiltonian.  (Indeed, the spin-triplet 
superconducting terms are certainly not 
gauge equivalent to the terms associated with $t_{\uparrow -\downarrow}$, guaranteeing
that the $SU(2)$ gauge 
symmetry is fully broken to $Z_2$, rather than to a residual $U(1)$ as might otherwise be the case).  
As usual the phase fluctuations $\theta$ 
can be absorbed, by means of a gauge transformation, into the longitudinal modes of the broken $U
(1)$ generator.

{\it As an aside: } (Eq. (\ref{Eq_Flucts}) reveals that the longitudinal modes of the broken generators are confined to $x-y$ chains and $z$ links in the lattice respectively.  Since the corresponding 
gauge fluctuations are no 
longer purely transverse in the condensed phase, this means that only the residual $Z_2$ gauge field 
and the amplitude fluctuations 
are free to propagate in both dimensions of the lattice.  This explains, to a large degree, why the effect of including these bosons in the theory is so innocuous.)

%{\it Dynamical fluctuations in the amplitudes of the fields $\Theta$ and $\Phi$? }

In summary, the fluctuations about mean-field are described by the real scalars $\rho, \tilde{\rho},$ and $ \phi$, describing fluctuations in the amplitudes of the various condensed bosonic fields, and the $SU(2)$ gauge field which is higgsed in a bi-adjoint representation to a residual symmetry group $Z_2$, which we may consider to have absorbed the remaining phase fluctuations as two Goldstone bosons.

\section{Mean-field theory of the gapped B phase} \label{BMFSect}

Here we describe the mean-field theory in the presence of the $3$-spin interaction which leads to the gapped topological B phase.  We will show that the band structure discussed in Sect. \ref{BGapSect} is, up to irrelevant operators, a saddle-point of an appropriate action, and thus constitutes at least a self-consistent mean-field solution to the fermion problem, if not a global minimum of the action.

We begin by re-writing the $3$-spin interaction as a sum of products of $6$-fermion interaction terms:
\ba
S^x_i S^y_j S^z_k &=& \frac{i}{8} \left (  f^\dag_{i \uparrow} f^\dag_{j \uparrow} f_{j \downarrow} f_{i \downarrow}  
- f_{i \uparrow} f_{j \uparrow} f^\dag_{j \downarrow} f^\dag_{i \downarrow}  \right . \\
&&\left. 
+ f^\dag_{i \uparrow} f_{j \uparrow} f^\dag_{j \downarrow} f_{i \downarrow}  
-f^\dag_{j \uparrow} f_{i \uparrow} f^\dag_{i \downarrow} f_{j \downarrow} \right )  
 \left( 2 f^\dag_{k \downarrow} f_{k \downarrow} - 1\right ) \nonumber
\ea
where we have used $n_{i \uparrow} = 1 - n_{i \downarrow}$ to express $S_z$ in terms of down spins only.  Of the possible fermion bilinears, only $(f^\dag_{i \uparrow} f_{j \uparrow}), ( f^\dag_{i \downarrow}  f_{j \downarrow})$, and $( f^\dag_{j \downarrow}  f_{k \downarrow})$ (together with their analogues in the particle-particle and hole-hole channels) have non-vanishing expectation values at mean-field.  (  $\langle  f^\dag_{k \downarrow}  f_{k \downarrow}-  f^\dag_{k \uparrow}  f_{k \uparrow} \rangle =0$). This gives us two possible ways to replace two of the $3$ fermion bilinears by their mean-field values.  First, we may take:
\ba \label{Eq_A2}
 \frac{i}{8} \left (\langle  f^\dag_{i \uparrow} f^\dag_{j \uparrow} \rangle \langle f_{j \downarrow} f_{i \downarrow}  \rangle
-\langle f_{i \uparrow} f_{j \uparrow}\rangle \langle f^\dag_{j \downarrow} f^\dag_{i \downarrow}\rangle  \ \ \ \ \ \ \ \ \ \ \ \  \right .  \\
\left. 
+\langle f^\dag_{i \uparrow} f_{j \uparrow}\rangle\langle f^\dag_{j \downarrow} f_{i \downarrow}  \rangle
-\langle f^\dag_{j \uparrow} f_{i \uparrow}\rangle \langle f^\dag_{i \downarrow} f_{j \downarrow} \rangle \right )  
 \left( 2 f^\dag_{k \downarrow} f_{k \downarrow} - 1\right ) \nonumber
 \ea
which vanishes in the mean-field solution relevant to the Kitaev model as the fermion bilinears are purely imaginary in position space.   The only remaining possibility is:
\ba
 \frac{i}{8 } \left [ \langle f^\dag_{i \uparrow}  f^\dag_{j \uparrow} \rangle \langle  f^\dag_{k \downarrow} f_{j \downarrow}  \rangle  f_{i \downarrow} f_{k \downarrow} -   \langle f^\dag_{i \uparrow}  f^\dag_{j \uparrow} \rangle \langle f_{j \downarrow}  f_{k \downarrow} \rangle  f^\dag_{k \downarrow} f_{i \downarrow}  \right . \n
  \left . - \langle f_{i \uparrow}  f_{j \uparrow} \rangle \langle f^\dag_{k \downarrow}  f^\dag_{j \downarrow} \rangle  f^\dag_{i \downarrow} f_{k \downarrow} +  
   \langle f_{i \uparrow}  f_{j \uparrow} \rangle \langle f^\dag_{j \downarrow}  f_{k \downarrow} \rangle  f^\dag_{k \downarrow} f^\dag_{i \downarrow}  \right . \n
   \left .+ \langle f^\dag_{i \uparrow}  f_{j \uparrow} \rangle \langle f^\dag_{k \downarrow}  f^\dag_{j \downarrow} \rangle  f_{i \downarrow} f_{k \downarrow} -  
   \langle f^\dag_{i \uparrow}  f_{j \uparrow} \rangle \langle f^\dag_{j \downarrow}  f_{k \downarrow} \rangle  f^\dag_{k \downarrow} f_{i \downarrow}  \right . \n
    \left . +  \langle f^\dag_{j \uparrow}  f_{i \uparrow} \rangle \langle f^\dag_{k \downarrow} f_{j \downarrow}   \rangle f^\dag_{i \downarrow} f_{k \downarrow}  -   \langle f^\dag_{j \uparrow}  f_{i \uparrow} \rangle \langle f_{j \downarrow}  f_{k \downarrow} \rangle  f^\dag_{k \downarrow} f^\dag_{i \downarrow}  \right ]      
 \ea
Taking $<ij>$ to be an $x$-link and $<jk>$ to be a $z$-link, and substituting in the mean-field values given in Eq. (\ref{Eq_MFUp}), this becomes:
\ba
&& \frac{i}{2^7 }   \left [  f_{i \downarrow} f_{k \downarrow} - f^\dag_{k \downarrow} f^\dag_{i \downarrow} +   f^\dag_{k \downarrow} f_{i \downarrow} - f^\dag_{i \downarrow} f_{k \downarrow}   \right ]      \n
 &=&  \frac{i}{2^7 }   \left ( f^\dag_{i\downarrow} - f_{i\downarrow} \right )
 \left( f^\dag_{k\downarrow} -f_{k\downarrow} \right )
 \ea
In light of the correspondence (\ref{Eq_Mar2Dir}) between our Dirac fermions and the Majorana basis originally used to diagonalize the problem, this is exactly the term originally proposed by Ref. \onlinecite{Kitaev06a} to
break $\mathbf{\hat{T}}$ and open a gap in the B phase.  

Before analyzing the resulting band structure, let us understand why we may simply replace the fermion bilinears by their mean-field values, as we have blithely done above.  
In fact, we can modify the Lagrangian (\ref{Eq_HSDec}) to produce just such a term at mean-field level.  To see why this is so, we consider the action:
\be \label{Eq_LFermi}
\mac{L}_{\mac{F}} =  \chi_1^\dag \chi_1 + \chi_2^\dag \chi_2 + i J' \chi_1^\dag \chi_2^\dag \chi^\dag_3 + h.c. \ \ \ .
\ee
We will show that $\mac{L}_{\mac{F}} $ is well-approximated by the Hubbard-Stratonovich-like action:
\ba \label{Eq_Test}
\mac{L} &=& - |\Phi_1|^2- |\Phi_2|^2 + \chi_1 \Phi_1  + \chi_2 \Phi_2 \n
&&- i J' \left(  \Phi_1 \Phi_2  - \chi_2^\dag \Phi_1 - \chi_1^\dag \Phi_2 \right)\chi^\dag_3 + h.c.
\ea
where $\chi_{1,2,3}$ are fermion bilinears.  (The Lagrangian (\ref{Eq_HSDec}) is of the general form of the quadratic terms in Eq. (\ref{Eq_Test}), albeit with more different scalar fields.  This multiplicity of indices will not affect our qualitative result).
The saddle-point equations are:
\ba
\Phi_1 &=& \chi_1^\dag - i J'\chi_3 ( \Phi^\dag_2 - \chi_2)  \n
   \Phi_2 &=& \chi_2^\dag  - i J'\chi_3  ( \Phi^\dag_1 - \chi_1)  
\ea
For $J'=0$, the  saddle-point equations specify that $\Phi_i = \chi_i^\dag$.  This is also the unique solution of the saddle-point equations for $J' \neq 0$ (though in this case one might worry about instabilities which tend to drive $\Phi_{1,2}$ towards $\infty$ if $\langle \chi_3 \rangle \neq 0$).  Hence the extra term does not modify the structure of the mean-field equations, except inasmuch as $\langle \chi_{1,2} \rangle$ might be modified by the new interaction.

As in the standard Hubbard-Stratonovich decoupling, we would like to integrate out $\Phi_{1,2}$ to obtain $\mac{L}_{\mac{F}} $.    As the Lagrangian (\ref{Eq_Test}) is no longer quadratic in the variables $\Phi_i, \chi_i$, we will not be able to perform the integral exactly; rather, we will obtain $\mac{L}_{\mac{F}}$
as the lowest-order term in an expansion in $J'$.  To see this, it is helpful to re-express $\mac{L}$ as:
\ba
\mac{L} %&=& | \Phi_1 - \chi_1^\dag |^2 + | \Phi_2 - \chi_2^\dag |^2 - iJ' (\Phi_1 - \chi_1^\dag) (\Phi_2 - \chi_2^\dag) \chi^\dag_3 + h.c. + \mac{L}_{\mac{F}} \n
%&& + \chi_1^\dag \chi_1 + \chi_2^\dag \chi_2 + J' \chi_1^\dag \chi_2^\dag \chi^\dag_3 + h.c. \n
 &=& - | \tilde{\Phi}_1  |^2 - |  \tilde{\Phi}_2  |^2 - i J'  \tilde{\Phi}_1  \tilde{\Phi}_2  \chi^\dag_3 + h.c. + \mac{L}_{\mac{F}}
 \ea
where $ \tilde{\Phi}_i \equiv \Phi_i - \chi^\dag_i$.  In the standard Hubbard-Stratonovich transformation there would be at this point no cross-terms coupling fermions to the scalar fields.  We could therefore integrate out the latter exactly and this prove that (\ref{Eq_Test}) is exactly equivalent to $\mac{L}_{\mac{F}}$.  Here we are unable to eliminate the cross-term $\Phi_1 \Phi_2 \chi^\dag_3$ by further shifting the scalar fields, so that integrating out the $\tilde{\Phi}$ fields will not reproduce $\mac{L}_{\mac{F}}$ exactly.  If we take $J'$ small, however, we may consider the effect of the cross-term perturbatively, and ask what the undesired additions to the fermionic action will be.   The exact correction is given by evaluating  the series:
\ba \label{Eq_LCorrect}
\delta \mac{L}_{\mac{F}} &=&  \log \left \{ \int \left[ D \tilde{\Phi}_1 \right] \left[ D \tilde{\Phi}_2 \right] e^{ i \int  | \tilde{\Phi}_1  |^2 + |  \tilde{\Phi}_2  |^2 } \right . \n
&& \left. \sum_{n=0}^{\infty} \frac{(i J')^n}{n!}  \left( \tilde{\Phi}_1  \tilde{\Phi}_2  \chi^\dag_3 + h.c. \right )^n \right \} \ \ \ .
\ea

Terms with $n$ odd integrate to $0$ since the action contains only even powers of $\tilde{\Phi}_{i}$.  Hence the leading correction is of order $J'^2$; to linear order in $J'$, then, we have recovered exactly the fermionic action we wanted.  Since the scalar-scalar-fermion bilinear interaction is decidedly irrelevant (all scalars here are massive), we may conclude that the difference between the action (\ref{Eq_Test}) and the true fermionic action $\mac{L}_{\mac{F} }$ is unimportant at least for the low-energy physics. 

The general form of this correction is simple to understand.  The leading-order correction in the series (\ref{Eq_LCorrect}) is proportional to  $ \frac{ (J')^2}{2} \chi^\dag_3 \chi_3 $.   If we take $\chi_3$ to have the form $f_{i \downarrow} f_{k \downarrow}$, then we have $\chi^\dag_3 \chi_3 =( \chi^\dag_3 \chi_3)^r = \hat{n}_{i \downarrow}  \hat{n}_{k \downarrow} $ for all $r$, and all terms in the series induce the same type of `extraneous' interaction,  which is to induce a second-neighbor `Coulomb repulsion' term.   

We conclude that at least the low-energy structure of the phase we are interested in can be obtained by studying the Lagrangian (\ref{Eq_Test}).  We may now proceed as in Sect. \ref{MFSect}, obtaining a mean-field solution which satisfies:
\be
\langle \Phi_i \rangle=\langle \chi_i^\dag \rangle
\ee
As noted above, the mean-field consistency conditions are identical to those at $J' =0$; the only new feature of this saddle point is that it now includes quadratic terms coupling fermions on the same sublattice, such as:
\be
J\ \langle  \Phi_1 \rangle \langle \Phi_2 \rangle f^\dag_{i \downarrow} f^\dag_{k \downarrow} \ \ \ .
\ee
This means that, to lowest order in $J'$, the effect of the $3$-spin interaction is, exactly as originally postulated by Ref. \onlinecite{Kitaev06a}, to modify the band structure by adding next-nearest neighbor quadratic couplings.  (We now also have to contend with the $4$ fermion interactions; however, when the quadratic problem has no Fermi surface, we do not expect these to be associated with instabilities of the free fermion problem and hence we can safely drop them without altering the qualitative nature of the physics.) 

\subsection{Form of the mean-field Hamiltonian with $3$ spin interactions}

Here we will derive the expression (\ref{Eq_HMF2}) for the terms induced by the set of all $3$-spin interactions at mean-field.
There are three distinct $3$-spin interactions that we must consider: 
\ba 
S^x_i S^y_j S^z_k    &\ \ \ & \mbox{if } r_{ik} = \hat{l}_1 \n
 S^y_i S^x_j S^z_k    &\ \ \ & \mbox{if } r_{ik} = \hat{l}_2 \n
 S^x_i S^z_j S^y_k    &\ \ \ & \mbox{if } r_{ik} = \hat{x} \ \ \ .
 \ea
The contributions to mean-field involve decoupling the resulting $6$-fermion interactions into combinations of a pair of $2$-point functions multiplying a fermion bilinear.  

First, we show that only contributions multiplying bilinears of the form $f_{i \sigma} f_{k \sigma}, f^\dag_{i\sigma} f_{k\sigma}$, etc., are non-vanishing.  The mean-field eigenfunctions imply that  $\langle S^{\sigma}_i \rangle =0$ on each site.  To show that  $\langle S^{\sigma}_i S^{\sigma'}_j \rangle =0$ if $\sigma \neq \sigma'$, we first note that if $\sigma = x,y$ and $ \sigma'=z$, any grouping of the resulting $4$-fermion interaction into pairs involves one term in each pair which contains both a spin up and spin down fermion.  Since the $2$-point functions of all terms involving spin flips are strictly $0$, these terms consequently all vanish.  If $\sigma = x, \sigma'=y$, then we have:
\ba
- i ( f^\dag_{i \uparrow } f_{i \downarrow} + f^\dag_{i \downarrow} f_{i \uparrow} )( f^\dag_{j \uparrow } f_{j \downarrow} - f^\dag_{j \downarrow} f_{j \uparrow} ) \n
= i \left(  \langle f^\dag_{i \uparrow}   f^\dag_{j \uparrow}   \rangle  \langle  f_{i \downarrow}  f_{j \downarrow} \rangle  + \langle   f^\dag_{i \downarrow}  f_{j \downarrow}\rangle  \langle f^\dag_{j \uparrow }  f_{i \uparrow }  \rangle  - h.c \right ) 
\ea
which vanishes since the $2$-point function on every link is purely imaginary, so that the products shown are purely real.  

The only remaining possibility is terms in which the $2$-point functions whose mean-field expectation we take involve fermion operators from all $3$ sites.  Since all $2$-point functions between sites $i$ and $k$ vanish at mean-field (this is guaranteed by the discrete symmetries $\mathbf{\hat{C}}$ and  $\mathbf{\hat{T}}$), the only posibility is terms which multiply fermion bilinears which couple the sites $i$ and $k$.  

Our next task is to understand the precise form of these terms.  For $r_{ij} = \hat{l}_{1,2}$, it is convenient to write $S^z_i = f_{i \downarrow}f^\dag_{i \downarrow} - f^\dag_{i \downarrow} f_{i \downarrow}$; for $r_{ij} = \hat{x}$, we write  $S^z_i =  f^\dag_{i \uparrow} f_{i \uparrow} -f_{i \uparrow}f^\dag_{i \uparrow} $.  The resulting expressions contain couplings only between the spin-down fermions on sites $i$ and $k$.   Thus the $3$-spin interaction does not modify the band structure of the spin-up fermions, which remain localized, at least at the mean-field level. 
% {\it The $4$-spin interactions may have an effect here?}  

\begin{widetext}
The quadratic couplings between the down spins induced by the $3$-spin interactions can be expressed:
\ba
S^x_i S^y_j S^z_k    &= &  \frac{i}{8} \left[  ( T^{(1)}_{ijk; \downarrow} +  T^{(3)}_{ijk; \downarrow} ) f^\dag_{k \downarrow}   f_{i \downarrow} +( T^{(2)}_{ijk; \downarrow} +  T^{(4)}_{ijk; \downarrow}  )f_{k \downarrow}   f_{i \downarrow}  
% \right . \n  && \left.
+ (T^{(6)}_{ijk; \downarrow}+ T^{(8)}_{ijk; \downarrow} )  f^\dag_{i \downarrow}   f_{k \downarrow}  + (T^{(5)}_{ijk; \downarrow}+ T^{(7)}_{ijk; \downarrow} )  f^\dag_{i \downarrow}   f^\dag_{k \downarrow} \right ] \n
S^y_i S^x_j S^z_k    &= &  \frac{i}{8} \left[  ( T^{(1)}_{ijk; \downarrow} -  T^{(3)}_{ijk; \downarrow} ) f^\dag_{k \downarrow}   f_{i \downarrow} +( T^{(2)}_{ijk; \downarrow} - T^{(4)}_{ijk; \downarrow}  )f_{k \downarrow}   f_{i \downarrow}  
%\right . \n  && \left.
- ( T^{(6)}_{ijk; \downarrow}- T^{(8)}_{ijk; \downarrow} )  f^\dag_{i \downarrow}   f_{k \downarrow}  - ( T^{(5)}_{ijk; \downarrow}- T^{(7)}_{ijk; \downarrow} )  f^\dag_{i \downarrow}   f^\dag_{k \downarrow} \right ] \n
S^x_i S^z_j S^y_k    &= &  \frac{i}{8} \left[  ( T^{(1)}_{ijk; \uparrow} -  T^{(3)}_{ijk; \uparrow} ) f^\dag_{k \downarrow}   f_{i \downarrow} +( T^{(2)}_{ijk; \uparrow} - T^{(4)}_{ijk; \uparrow}  )f_{k \downarrow}   f_{i \downarrow} 
% \right . \n && \left.
- ( T^{(6)}_{ijk; \uparrow}- T^{(8)}_{ijk; \uparrow} )  f^\dag_{i \downarrow}   f_{k \downarrow}  - ( T^{(5)}_{ijk; \uparrow}- T^{(7)}_{ijk; \uparrow} )  f^\dag_{i \downarrow}   f^\dag_{k \downarrow} \right ] \n
\ea
\end{widetext}
with
\ba
 T^{(1)}_{ijk; \sigma} =& - \langle f^\dag_{i \uparrow}  f^\dag_{j \uparrow} \rangle \langle  f_{j  \sigma}f_{k  \sigma}  \rangle \ \ \ &= -\frac{16}{J_{ij} J_{jk} } \left( \Delta^{(ij)}_{\uparrow} \right)^*  \Delta^{(kj)}_{ \sigma} \n
 T^{(2)}_{ijk; \sigma} =&- \langle f^\dag_{i \uparrow}  f^\dag_{j \uparrow} \rangle \langle  f^\dag_{k  \sigma} f_{j  \sigma}  \rangle \ \ \ &= -\frac{16}{J_{ij} J_{jk} } \left( \Delta^{(ij)}_{\uparrow} \right)^* \left( t^{(kj)}_{ \sigma}  \right)^* \n
 T^{(3)}_{ijk; \sigma} =& - \langle f^\dag_{i \uparrow}  f_{j \uparrow} \rangle \langle f^\dag_{j  \sigma}  f_{k  \sigma} \rangle \ \ \ &=-\frac{16}{J_{ij} J_{jk} }  \left( t^{(ij)}_{\uparrow} \right)^*t^{(kj)}_{ \sigma}\n
T^{(4)}_{ijk;  \sigma} =&  \langle f^\dag_{i \uparrow}  f_{j \uparrow} \rangle \langle f^\dag_{k  \sigma}  f^\dag_{j  \sigma}  \rangle \ \ \ &= \frac{16}{J_{ij} J_{jk} } \left(t^{(ij)}_{\uparrow} \right)^*  \left ( \Delta^{(kj)}_{ \sigma} \right)^*\n
T^{(5)}_{ijk;  \sigma} =&  \langle f^\dag_{j \uparrow}  f_{i \uparrow} \rangle \langle f_{j  \sigma}  f_{k  \sigma}  \rangle  \ \ \ & =-  \frac{16}{J_{ij} J_{jk} } t^{(ij)}_{\uparrow} \Delta^{(kj)}_{ \sigma}   \n
 T^{(6)}_{ijk;  \sigma} =& \langle f^\dag_{j \uparrow}  f_{i \uparrow} \rangle \langle f^\dag_{k  \sigma} f_{j  \sigma}   \rangle \ \ \ &=\frac{16}{J_{ij} J_{jk} }  t^{(ij)}_{\uparrow}   \left( t^{(kj)}_{ \sigma}  \right)^* \n
 T^{(7)}_{ijk;  \sigma} =& \langle f_{j \uparrow}  f_{i \uparrow} \rangle \langle f^\dag_{j  \sigma}  f_{k  \sigma} \rangle  \ \ \ &= \frac{16}{J_{ij} J_{jk} }  \Delta^{(ij)}_{\uparrow}  t^{(k j)}_{ \sigma}\n
T^{(8)}_{ijk;  \sigma} =&   \langle f_{j \uparrow}  f_{i \uparrow} \rangle \langle f^\dag_{k  \sigma} f^\dag_{j  \sigma}  \rangle  \ \ \ &=   \frac{16}{J_{ij} J_{jk} } \Delta^{(ij)}_{\uparrow}  \left( \Delta^{(kj)}_{ \sigma}  \right ) ^* \n
\ea
where we have used $t^{(jk) *} = t^{(kj)}$, $\Delta^{(jk) *} = \Delta^{(kj)}$.  
(Here we have defined $\Delta^{(ab)} = \Delta^{(x,z)}$ on $x$ and $z$ links, and $- \Delta^{(y)}$ on $y$ links, in accordance with Eq. (\ref{Eq_MFUp}) ).

We next substitute in the mean-field values given in Eq. (\ref{Eq_MFUp}) for $t, \Delta$ on each link.  We take $t$ to be the hopping from sublattice $1$ to sublattice $2$ ( $t^{(ij)}_{\sigma} = \langle f^\dag_{ \vec{R} 1 \sigma}f_{\vec{R'} 2 \sigma} \rangle $), and similarly for $\Delta$.  Here we write the induced quadratic couplings between two sites on sublattice $1$;  the couplings between sites on sublattice $2$ are the same, but with $r_{ij} \rightarrow - r_{ij} $.   

 For $r_{ij} = \hat{l}_1$, the interaction is of the form $J' S^x_i S^y_j S^z_k $, with $ij$ an $x$-link and $jk$ a $z$-link.  We thus have $\Delta^{(jk)}_{ \downarrow} = 0$, giving an interaction of:
\ba
 2 i J' \left[  - t^{(x)*}_{\uparrow}   t^{(z)}_{\downarrow}  f^\dag_{k \downarrow}   f_{i \downarrow} - \Delta^{(x)*}_{\uparrow}  t^{(z)*}_{ \downarrow} f_{k \downarrow}   f_{i \downarrow}  \right. \n
 \left. 
 + t^{(x)}_{\uparrow}  t^{(z)*}_{ \downarrow}   f^\dag_{i \downarrow}   f_{k \downarrow}  + \Delta^{(x)}_{\uparrow} t^{(z)}_{\downarrow}   f^\dag_{i \downarrow}   f^\dag_{k \downarrow} \right ] \ \ \ 
\ea
with
\ba 
\Delta^{(x)}_{\uparrow}  =&- i \frac{J_x}{16}  \ \ \      
 t^{(x)}_{\uparrow}     =& - i \frac{J_x}{16}      \ \ \           t^{(z)}_{\downarrow}  =  -i \frac{J_z}{8} \ \ \ .
\ea
Similarly, for $r_{ij} = \hat{l}_2$, we have $J' S^y_i S^x_j S^z_k $, with $ij$ a $y$-link and $jk$ a $z$-link.  Hence
again $   \Delta^{(jk)}_{ \downarrow} =0 $, and 
the interaction is:
\ba
 2 i J' \left[   t^{(y)*}_{\uparrow}  t^{(z)}_{ \downarrow}  f^\dag_{k \downarrow}   f_{i \downarrow} + \Delta^{(y)* }_{\uparrow} t^{(z)* }_{ \downarrow} f_{k \downarrow}   f_{i \downarrow}  \right. \n
 \left. 
 - t^{(y)}_{\uparrow}  t^{(z)*}_{ \downarrow} f^\dag_{i \downarrow}   f_{k \downarrow}  - \Delta^{(y)}_{\uparrow} t^{(z)}_{\downarrow}   f^\dag_{i \downarrow}   f^\dag_{k \downarrow} \right ] \  \ \ 
\ea
with
\ba 
\Delta^{(y)}_{\uparrow}  =& i \frac{J_y}{16}  \ \ \ \ \     
 t^{(y)}_{\uparrow}     =& - i \frac{J_y}{16}          \ \ \ \ \         t^{(z)}_{\downarrow}  = - i \frac{J_z}{8} \ \ \ .
\ea
For $r_{ij} = \hat{x}$, we have $J' S^x_i S^z_j S^y_k $, with $ij$ an $x$-link and $jk$ a $y$-link.  This gives the interaction:
\ba
&  i J'  \left[   \left( \Delta^{(x)*}_{\uparrow} \Delta^{(y)*}_{  \uparrow} 
   -   t^{(x)*}_{\uparrow}  t^{(y)}_{  \uparrow}  \right) f^\dag_{k \downarrow}   f_{i \downarrow} +\left( \Delta^{(x)*}_{\uparrow} t^{(y)*}_{  \uparrow}  \right .  \right . \n 
& \left.
 \left.+ t^{(x)*}_{\uparrow}  \Delta^{(y)*}_{  \uparrow} \right)  f_{k \downarrow}   f_{i \downarrow}  
+ \left(- t^{(x)}_{\uparrow}   t^{(y)*}_{  \uparrow} - \Delta^{(x)}_{\uparrow}  \Delta^{(y)}_{  \uparrow}   \right )  f^\dag_{i \downarrow}   f_{k \downarrow}  \right . \n 
 & \left. + \left(t^{(x)}_{\uparrow}    \Delta^{(y)*}_{  \uparrow} + \Delta^{(x)}_{\uparrow}  t^{(y)}_{  \uparrow}  \right )  f^\dag_{i \downarrow}   f^\dag_{k \downarrow} \right ]  \ \  \ \ \ \ \ \ \ \ \ \ 
\ea
with
\ba 
     \Delta^{(x)}_{ \uparrow} =& - i \frac{J_x}{16}  \ \ \ \ \      \Delta^{(y)}_{\uparrow}  =& i \frac{J_y}{16}   \n
 t^{(x)}_{\uparrow}  =&  -i \frac{J_x}{16}      \ \ \ \ \            t^{(y)}_{\uparrow}     =& - i \frac{J_y}{16} 
\ea
In all $3$ cases, we obtain the mean-field interaction:
\ba
&& \pm 2 i J'  \left[    f^\dag_{k \downarrow}   f_{i \downarrow} -  f_{k \downarrow}   f_{i \downarrow} 
 + f^\dag_{i \downarrow}   f_{k \downarrow}  -  f^\dag_{i \downarrow}   f^\dag_{k \downarrow} \right ] 
\n
&=&\pm 2 i J'  \left ( f^\dag_{k \downarrow}  - f_{k \downarrow}  \right ) \left ( f^\dag_{i \downarrow}  - f_{i \downarrow}  \right ) \ \ \ .
\ea
We see that this induces a coupling only between Majorana modes in the dispersing band, leaving the band structure of the Majoranas localized on the $z$-links unaltered.  

Hence, the net effect of adding the $3$-spin interaction, at mean-field level, is exactly to add the next-nearest neighbor couplings to the dynamical Majorana modes, while leaving the localized modes unchanged.

 \section{Inducing Chern-Simons terms by integrating out fermions in the gapped B phase}  \label{FermiLoopSect}
 
 Here we will consider the $1$-loop perturbative correction to the effective $U(1)$ gauge field propagator due to the low-energy fermions in the gapped phase.  We demonstrate that though the Dirac point is intrinsically a property of the band structure of the superconductor -- such that the electron bubble has both particle- particle and particle -hole contributions -- the matrix structure about the Dirac point is such that integrating out the low-energy fermions produces exactly the same Chern-Simons correction to the effective action as doing so for a normal Dirac cone.  
 
Since the Dirac cone is in only one of the $4$ fermion bands, and we are interested only in the long-wavelength theory, we will isolate the effect of the propagator of the dispersing Majorana band.  The general form of the spin-down propagator in the gapped B phase is
\begin{widetext}
\ba
G_{\downarrow \downarrow q} &=& \frac{1}{2} \left \{ 
\frac{1}{4 m_q^2+\omega ^2+| \Delta_q-t_q|^2}   
\left(
\begin{array}{cccc}
 -2 m_q-i \omega  & -i (\Delta_q -t_q) & 2 m_q+i \omega  & i (\Delta_q -t_q)  \\
 i (\Delta^*_q-t^*_q) & 2 m_q-i \omega  & -i (\Delta^*_q-t^*_q) & i \omega -2 m_q  \\
 2 m_q+i \omega  & i (\Delta_q -t_q) & -2 m_q-i \omega  & -i (\Delta_q -t_q)  \\
 -i (\Delta^*_q-t^*_q) & i \omega -2 m_q & i (\Delta^*_q-t^*_q) & 2 m_q-i \omega 
\end{array} 
\right) \right  . \n
&& + \left .  
\frac{1}{\omega ^2+| \Delta_q+t_q|^2}   \left( 
\begin{array}{cccc} 
 -i \omega  & i (\Delta_q +t_q) & -i \omega  & i (\Delta_q +t_q) \\ 
 -i (\Delta^*_q+t^*_q) & -i \omega  & -i (\Delta^*_q+t^*_q) & -i \omega  \\ 
 -i \omega  & i (\Delta_q +t_q) & -i \omega  & i (\Delta_q +t_q)  \\
 -i (\Delta^*_q+t^*_q) & -i \omega  & -i (\Delta^*_q+t^*_q) & -i \omega  
\end{array}
\right) \right \}
\ea
\end{widetext}
where we use the basis $ \psi = \left ( \begin{array}{cccc} c_{q1} &c_{q2}  &c^\dag_{-q 1} &c^\dag_{-q 2} \end{array} \right ) ^T$.

Here we choose $t_q = -2 J_z - J_x e^{i \vec{q} \cdot \hat{l}_1} - J_y e^{i \vec{q} \cdot \hat{l}_2} , \Delta_q = J_x e^{i \vec{q} \cdot \hat{l}_1} + J_y e^{i \vec{q} \cdot \hat{l}_2}$.  In this case the bottom line is the propagator of the flat band (energies given by $\pm | t + \Delta| = \pm 2 J_z$; the top line is the propagator of the dispersing band, which captures all of the low-energy physics near the Dirac cones.  It is easy to check that cross-terms between the two spin down bands vanish at $1$-loop order in the fermion correction, so that we will drop contributions of the flat gapped band entirely.

In the vicinity of the Dirac cone $\vec{q} = (\frac{ 4 \pi}{3}, 0)$, at the isotropic point $J_x= J_y =J_z$, we have 
\be 
\Delta_q - t_q  \approx  \sqrt{3} J  \ \ \ \ \ \ \ \ \ \ 
m_q \approx  \frac{3}{2} \sqrt{3} J'  \ \ \ .
\ee
Near this point in the Brillouin zone, then, the part of the propagator that we are interested in can be expressed as:
\ba \label{Eq_FProp2}
G_{c; q,\omega} &=& \frac{1}{2} \left( G_{c; q,\omega}^{(0)} + G_{c; q,\omega}^{(sc)} \right )  \\
 G_{c; q,\omega}^{(0)} &=& \frac{1}{4 m_q^2+\omega ^2+| \Delta_q-t_q|^2}   \left ( p^\mu \sigma_\mu + 2 m \sigma_z   \right ) \otimes \mathbf{1}  \n
 G_{c; q,\omega}^{(sc)} &=&\frac{1}{4 m_q^2+\omega ^2+| \Delta_q-t_q|^2}   \left ( p^\mu \sigma_\mu +2  m \sigma_z \right )  \otimes \sigma_x \nonumber 
\ea
with $\sigma^\mu = \left( \begin{array}{ccc} \mathbf{1} & \sigma_y & \sigma_x \end{array} \right )$.  
In addition to the usual term ($ G_{c; q,\omega}^{(0)} $), the fermion propagator contains an anomalous term ($G_{c; q,\omega}^{(sc)}$) due to the presence of superconductivity.  The  $2 \times 2$ matrix structure of both of these terms is, however, the same.

In this long-wavelength limit, the interaction between fermions and the gauge field is
\be
A^\mu_q \sum_k  \psi^\dag_k \gamma_\mu \psi_{k-q} - 2 \delta_{\mu 0} \delta_{q 0} 
\ee
where $\gamma_\mu = \sigma_\mu \otimes \mathbf{1}$, and the last term occurs due to normal ordering.  
(Here it should be understood that the sum encompasses only half the Brillouin zone).
The $1$-loop correction to the gauge field effective action induced by the fermion terms is therefore:
\be
\mac{L}^{(G)}_{\mu \nu} (\vec{p}, \Omega)= \int \frac{d^3 p }{ \left( 2 \pi\right ) ^3} Tr \left [ \gamma_\mu G_{c; q, \omega} \gamma_\nu G_{c; q+ p, \omega+ \Omega} \right ]
\ee

Using the expression (\ref{Eq_FProp2}), we find that
traces of the cross-terms between $G_{c; q,\omega}^{(0)}$ and $G_{c; q,\omega}^{(sc)}$ vanish, leaving:
\ba
\mac{L}^{(G)}_{\mu \nu} (\vec{p}, \Omega)&=& \frac{1}{4} \left \{ 2 \mac{L}^{(1)}_{\mu \nu}  \right. \\
&& \left. +\int \frac{d^3 p }{ \left( 2 \pi\right ) ^3} Tr \left [ \gamma_\mu G^{(sc)}_{c; q, \omega} \gamma_\nu G^{(sc)}_{c; q+ p, \omega+ \Omega} \right ] \right \} \nonumber 
\ea
where $\mac{L}^{(1)}_{\mu \nu}$ is the effective action induced by the usual $2+1$ dimensional Dirac cone (appearing here with a multiplicative factor of $2$ since we have counted both terms of the form $f^\dag_{q i} f_{q i}$ and $f^\dag_{-q, i} f_{-qi}$, effectively counting the contribution of both Dirac cones).  The second contribution, due to the superconducting terms, also has precisely the same form as the first, since $G^{(sc)}$ has the same $2 \times 2$ structure as $ G^{(0)}$.  The factor of $\frac{1}{4}$ (due to the $\frac{1}{2}$ in  $G^{(0)}$ relative to its usual value) is exactly cancelled by the factor of $4$ from these contributions.
This gives exactly the $1$-loop correction expected from a single Dirac cone in QED, albeit with a mass of $2 m$ rather than $m$.

\bibliography{KitaevBib}

\end{document}